\documentclass[aps,prb,reprint,twocolumn,amsmath,amssymb,superscriptaddress]{revtex4-1}
\usepackage[dvipdfmx]{graphicx}    % Include figure files
\usepackage{dcolumn}               % Align table columns on decimal point
\usepackage{bm}                    % bold math
\usepackage{multirow}
\usepackage{color}
\usepackage{threeparttable}
\usepackage{dcolumn}
\usepackage{sidecap}
\newcommand{\braket}[1]{\langle #1 \rangle}        % <   #   >
\begin{document}

\preprint{APS/123-QED}

%... title
\title{Linear-response-based DFT+U method\\for exploring half-metallic Co-based full Heusler alloys}

%... authors
\author{Kenji Nawa}
\email[E-mail address: ]{nawa.kenji@nims.go.jp}
\affiliation{Research Center for Magnetic and Spintronic Materials (CMSM), 
             National Institute for Materials Science (NIMS),
             1-2-1 Sengen, Tsukuba, Ibaraki 305-0047, Japan}

\author{Yoshio Miura}
\affiliation{Research Center for Magnetic and Spintronic Materials (CMSM), 
             National Institute for Materials Science (NIMS), 
             1-2-1 Sengen, Tsukuba, Ibaraki 305-0047, Japan}
\affiliation{Center for Materials research by Information Integration, 
             National Institute for Materials Science (NIMS), 
             1-2-1 Sengen, Tsukuba, Ibaraki 305-0047, Japan}
\affiliation{Center for Spintronics Research Network (CSRN), 
             Graduate School of Engineering Science, Osaka University, 
             1-3 Machikaneyama, Toyonaka, Osaka 560-8531, Japan}

\date{\today}

%-- abstract ---------------------------------------------------------------------------------------------
\begin{abstract}
The density functional theory (DFT)+U method based on the linear response (LR) theory was applied to 
investigate the electronic structures of Co-based ternary full Heusler alloy Co$_2Y$Si for exploring 
half-metallic (HM) ferromagnets with a wide HM gap.
The LR-based DFT+U calculations tend to obtain a reasonable correlation parameter for $Y$ site,
while the correlation of Co site misleads to the unphysical ground state due to the 
overestimated parameter value that arises from the delocalized electronic structure of Co.
Furthermore, 
we found that the HM gap of Co$_2$MnSi originates from Co $e_u$ orbital in the conduction state and Co-Mn 
hybridizing $t_{2g}$ orbital in the valence state around the Fermi energy.
This means that the HM gap is a tunable property by selecting the $Y$ element and/or mixing several elements
into the $Y$ site through $t_{2g}$ atomic-orbital coupling.
Our LR-based DFT+U method was extended to other ternary Co$_2Y$Si and quaternary Co$_2$($Y$,Mn)Si.
We found that Co$_2$(Ti$_{0.25}$,Mn$_{0.75}$)Si and Co$_2$(Fe$_{0.25}$,Mn$_{0.75}$)Si show 
HM nature, with the Fermi energy being at almost the center of the minority band gap, which leads to high thermal 
stability.
\end{abstract}

\maketitle

%-- introduction ----------------------------------------------------------------------------------------
\section{introduction}\label{sec:intr}

A key property in emerging field of spintronics is the so-called half-metallicity (HM);
the majority and minority states are completely spin-polarized at the Fermi level,
where a finite density of states (DOS) exists 
for majority spin and an energy band gap is opened for minority spin.
For example, the use of HM materials as ferromagnetic electrodes 
in magnetic tunnel junctions (MTJs) is a straightforward way to enhance tunneling magnetoresistance 
(TMR) ratio~\cite{pla-54-225-1975},
leading to high-performance spintronics applications such as non-volatile magnetic random
access memories and read-head of ultrahigh-density hard-disk drives.
The family of Co-based full Heusler alloys has received considerable attention, as some of these 
have a potential to possess a high spin polarization ($P$) or ultimately HM ($P=100$~\%)
in addition to a high Curie temperature, e.g., 985~K for Co$_2$MnSi~\cite{jmmm-38-1-1983} and 1100~K for 
Co$_2$FeSi~\cite{prb-72-184434-2005,apl-88-032503-2006}.

The spin polarization of electrodes in an MTJ device can be evaluated using the Julli\`{e}r 
model~\cite{pla-54-225-1975} with a simple formula
$\mathrm{TMR}=\frac{2 P_1 P_2}{1 - P_1 P_2} \times 100~(\%)$,
where $P_1$ and $P_2$ are the tunneling spin polarizations of two ferromagnetic electrodes in the MTJ.
For Co$_2$MnSi MTJ with an aluminum oxide (Al-O) barrier, Sakuraba {\it et al.} observed the spin polarization to be over 
$80~\%$.~\cite{jjap-44-L1100-2005,apl-88-192508-2006} 
Then, a high value of 95.4~\%, which may be close to a fully spin-polarized electronic structure, 
was reported for the MgO barrier MTJ~\cite{prb-94-094428-2016}.
However, $P_{1(2)}$ in the Julli\'{e}r formula is not the spin polarization
in the bulk system but the polarization of tunneling electrons in the MTJ.
The electronic structure of MTJ electrode differs from that of the bulk material because the band 
structure is drastically changed due to the interfacial effect arising from the insulating barrier.
The tunneling electrons are also influenced by spin-filtering effect.~\cite{prb-63-054416-2001}
These facts imply that there is difficulty in an accurate estimation of purely bulk spin polarization 
from the TMR of MTJ.

Point contact Andreev reflection (PCAR) technique has also been performed for spin polarization
in several Heusler alloys.
The conductance of metallic electrons is measured at cryogenic temperature to evaluate the spin polarization 
in PCAR; hence, $P_\mathrm{PCAR}$, referred to as PCAR-measured spin polarization, is expressed as
$P_\mathrm{PCAR}=
   \frac{\braket{N^\uparrow(E_\mathrm{F})v_\mathrm{F}^\uparrow} 
       - \braket{N^\downarrow(E_\mathrm{F})v_\mathrm{F}^\downarrow}}
        {\braket{N^\uparrow(E_\mathrm{F})v_\mathrm{F}^\uparrow}
       + \braket{N^\downarrow(E_\mathrm{F})v_\mathrm{F}^\downarrow}} \times 100~(\%)$.
Here, $N^\sigma(E_\mathrm{F})$ and $v^\sigma_\mathrm{F}$ are the DOS at Fermi energy and Fermi velocity with
spin index $\sigma$~($=\uparrow$ or $\downarrow$) in a diffusive regime~\cite{sci-282-85-1998,prb-75-054404-2007} 
where the current electrons are not assumed to be ballistic because of the mean-free path being shorter than the point 
contact size in actual experiments.
The $v^\sigma_\mathrm{F}$ is conductance of electrons, but the $d$ orbital 
localized around Fermi energy is not dominant in the current electron.
This indicates that the spin polarization originating from the $d$ electron is lost in the measured $P_\mathrm{PCAR}$.
Previous works reported that the current spin polarization deduced by PCAR is only 59~\% for
Co$_2$MnSi~\cite{jap-105-063916-2009}
and around 50~\% for Co$_2$FeSi~\cite{apl-89-082512-2006,jap-102-043903-2007,prb-87-220402R-2013}.
$P_\mathrm{PCAR}=64~\%$ is also observed in quaternary Co$_2$(Fe,Mn)Si.~\cite{prb-91-104408-2015}

Another critical subject to overcome is large temperature dependence of $P$.~\cite{jjap-44-L1100-2005,
apl-88-192508-2006,jmsj-38-45-2014,apl-93-112506-2008,prb-94-094428-2016,apl-89-082512-2006,
apl-101-252408-2012,prb-93-134403-2016}
Experimental studies have reported that although an extremely high value of TMR ratio is demonstrated at low temperature 
in the MTJs consisting of the Heusler electrodes and MgO barrier, 
a significant reduction in TMR at room temperature is observed,
for example, the TMR of 2010~\% at 4.2~K, but it decreases to only $335~\%$ at 290~K in 
Co$_2$MnSi/MgO/Co$_2$MnSi MTJ~\cite{prb-94-094428-2016} and 2610~\% at 4.2~K, but only 429~\% at 290~K in 
Co$_2$(Fe,Mn)Si/MgO/Co$_2$(Fe,Mn)Si\cite{jphysDapplphys-48-164001-2015}.
From the Julli\`{e}r model, the spin polarization $P=95~\%$ (98~\%) at low temperature decreases
to $P=79~\%$ (82~\%) at room temperature for MTJs with Co$_2$MnSi (Co$_2$(Fe,Mn)Si) electrode.
Similar situation occurred in a current-perpendicular-to-plane giant MR (CPP-GMR) device 
composed of Co$_2$(Fe,Mn)Si electrodes and nonmagnetic Ag spacer.~\cite{jmsj-38-45-2014,apl-101-252408-2012}
For explaining the strong thermal-dependence of TMR and GMR performances, it is known that spin-flip inelastic 
tunneling process induced by magnon excitation lowers $P$ in addition to
spin-conserving elastic tunneling at increased temperature.~\cite{apl-93-112506-2008,prb-94-094428-2016}
In this sense, a width of the energy band gap in the minority state is also important in the search for HM materials
to improve the weak resistivity with respect to temperature.

The {\it ab-initio} calculations based on DFT~\cite{pr-136-b864-1964,
pr-140-a1133-1965,rmp-71-1253-1999} are expected to play a leading role in the understanding of fundamental electronic 
and magnetic structures in material design using HM Heusler alloys.
In the framework of DFT calculation within local spin density approximation (LSDA),
Galanakis {\it et al.}~\cite{prb-66-174429-2002} presented an energy diagram of atomic orbital hybridization 
of Co$_2$MnGe system to clarify the mechanism of HM property;
the minority energy band gap at Fermi level originates from the $t_{1u}$ and $e_u$ orbitals, which are formed by  
the $d$ orbital hybridizations between two Co atoms sitting at different sublattices in a unit cell.
Numerous other studies have also been performed by DFT calculations.~\cite{prl-50-2024-1983,
jphysFmetphys-16-L211-1986,prb-53-1146-1996,jpsj-64-2152-1995,prb-66-094421-2002,jap-93-6844-2003,
prb-72-184415-2005}

However, a deal with correlation effects is a critical issue in the DFT study of a Heusler compound.
The standard DFT calculations based on mean-field approximations, such as LSDA and generalized gradient
approximation (GGA), often fail to predict the {\it true} ground-state electronic structures 
due to the presence of $d$ orbital localization in the vicinity of transition metal atoms, 
making the many-body effect problematic.
Various approaches introducing the many-body effect into the DFT scheme
have been proposed to recover the correlation problem being missed in LSDA and GGA; e.g.,
dynamical mean field theory (DMFT)~\cite{jphysCondMat-9-7359-1997,prb-57-6884-1998},
GW approximation~\cite{prb-78-155112-2008,prl-102-126403-2009}, and 
DFT+U method~\cite{prb-44-943-1991,jphysCondMat-9-767-1997}.
However, obtained electronic structures strongly depend on the employed method.
For example, in Co$_2$MnSi, the LDSA+DMFT calculations, where the dynamical correlation effect such as
the spin-flip term is considered quantitatively, were performed on the basis of the linear muffin-tin
orbital (LMTO) method~\cite{rmp-80-315-2008} and Korringa-Kohn-Rostoker (KKR) method~\cite{jphysDapplphys-42-084002-2009}.
The former indicates that the Fermi energy is found at the conduction edge of the minority state, 
while the latter is found at the valence edge.
The GW calculation~\cite{prb-86-245115-2012}, where the electronic self-energy correction
is included by the many-body perturbation theory, predicts that the Fermi energy lies between the valence and 
conduction bands of the minority state.
For these approaches, the huge computational cost is also a serious problem;
applying it to the MTJ model for properties including interfacial magnetocrystalline anisotropy and 
spin-dependent transport may be difficult.
On the other hand, 
the DFT+U method~\cite{prb-44-943-1991,jphysCondMat-9-767-1997}, in which parametrized on-site Coulomb ($U$) 
and exchange ($J$) interactions for $d$-orbital are introduced in the manner of the Hubbard 
model~\cite{ProcRSocA-285-542-1965,ProcRSocA-296-82-1967}, 
is a suitable approach on a practical level. 
Because of the efficient calculation cost, the DFT+U method can be applied to not only simple bulk materials 
but also large and realistic systems.

The suitable values of $U$ and $J$ for the DFT+U method are unknown;
they depend on the atomic species and surroundings of the atom.
A linear response (LR) approach~\cite{prb-71-035105-2005,prb-97-035117-2018} is an advanced way to 
determine the correlation terms theoretically and to exclude the {\it ad hoc} selection of the parameter values.
The +U values at respective localized atom sites can be evaluated using the response function of charge density 
obtained from the standard LSDA or GGA potential with low computational costs.
This method has been applied to various correlated systems
and succeeded in describing the ground state accurately.~\cite{prb-97-035117-2018,prb-71-035105-2005,
prb-70-235121-2004,prb-84-115108-2011,prb-89-085103-2014,
prb-92-075124-2015,jpcs-120-4919-2016,prb-94-035136-2016}
A recent study has also reported that the parameters are not transferable among 
different calculation methods due to non-negligible dependence on computational setups even in 
theoretically determined values.~\cite{prb-97-035117-2018}
This implies that the optimal correlation parameters for the system of interest must be estimated 
by the method used for the calculation; however, the application of this LR-based DFT+U method to Heusler compounds 
has been limited to structural phase 
transition~\cite{jphysCondMat-24-185501-2012}.

%... Fig.1
\begin{figure*}
\begin{center}
\includegraphics[width=1.9\columnwidth]{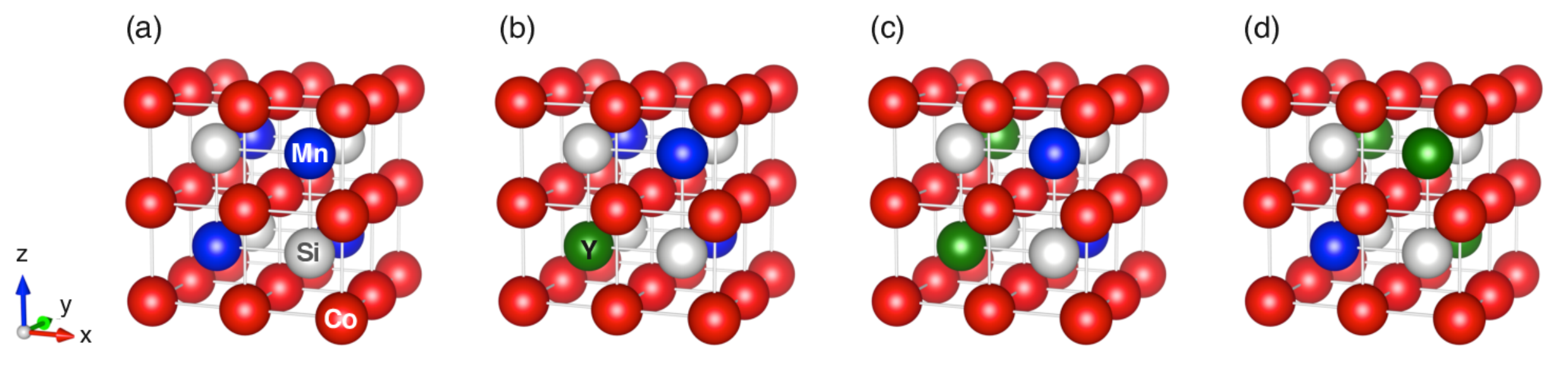}
\end{center}
\caption{(Color online)
         $L2_1$ symmetric crystal structures for 
         (a)~ternary Co$_2$MnSi,
         (b)~quaternary Co$_2$($Y$$_{0.25}$,Mn$_{0.75}$)Si, 
         (c)~Co$_2$($Y$$_{0.50}$,Mn$_{0.50}$)Si, and 
         (d)~Co$_2$($Y$$_{0.75}$,Mn$_{0.25}$)Si. 
         Red, blue, white, and green circles indicate Co, Mn, Si, and $Y$ atoms, respectively,
         where $Y$ is the $3d$ transition metal atom of Ti, V, Cr, or Fe.}
\label{fig_model}
\end{figure*}

In the present work, 
the electronic band calculations based on the DFT+U method are carried out for an $L2_1$ Co-based full Heusler
alloy to address these issues.
Focusing on the typical model Co$_2$MnSi, we argue the importance of correlation correction in the alloy
and the fundamental electronic structure for clarifying the origin of minority HM gap.
The LR calculations obtain a reasonable correlation parameter for the $Y$ site ($Y=$ Mn in Co$_2$MnSi) and this 
correction plays an important role for correlated electrons.
On the other hand, the correlation for Co site is unexpectedly overestimated, and thus, the obtained results 
are misled to a rather unphysical ground state.
The failure of an unreliable parameter of Co arises from the fact that the $3d$ electrons of Co site behave itinerary
in the alloy, which means that the mean-field approximations such as LSDA and GGA are enough to describe the 
electronic structure of Co site with high accuracy.
We also revealed an important $d$-orbital hybridization between Co and Mn that mainly dominates the minority HM gap.
The energy diagram proposed in this study suggests that the HM gap is tunable by a selected $Y$ atom and/or mixing
several elements into $Y$ site.
The results obtained from the LR-based DFT+U method, where the determined correlation parameter is 
incorporated into only the strongly correlated $Y$ site, are consistent with the experimental observations; 
moreover, this methodology is superior to the standard GGA calculation, especially in terms of electronic and magnetic 
properties.
This study is further extended to the other ternary Co$_2Y$Si and quaternary Co$_2$($Y_{x}$,Mn$_{1-x}$)Si, where 
a part of Mn is substituted with $3d$ transition metal $Y$ ($Y=$ Ti, V, Cr, and Fe) to explore the potential 
for the HM ferromagnet with a wide band gap.
The systematical calculations indicate that the ternary alloys are found to be ordinary ferromagnets, whose minority
bands do not have a finite gap at the Fermi energy, but quaternary Co$_2$(Ti,Mn)Si, Co$_2$(V,Mn)Si, and 
Co$_2$(Fe,Mn)Si alloys have a potential to be HM material if the composition of $Y$ is appropriately selected.

This paper is organized as follow.
In Sec.~\ref{sec:model-method}, the model and computational details are described, and the LR calculation 
procedures for the correlation parameters are overviewed.
Sec.~\ref{sec:CMS} revisits Co$_2$MnSi.
The effective on-site Coulomb interaction parameters for Co and Mn are first computed from the LR theory 
(Sec.~\ref{subsec:ulrt}).
The structural parameters, including equilibrium lattice constant and bulk modulus, are evaluated by standard 
GGA and GGA+U schemes with LR-determined parameters in Sec.~\ref{subsec:str}.
Using the obtained lattice constant, the electronic structures are investigated to clarify the HM origin within 
the GGA framework (Sec.~\ref{subsec:org-HM}).
The understanding of a fundamental band structure in GGA is essential for discussing the effects of 
correlation correction on Mn and Co, which is given in Sec.~\ref{subsec:effect-of-U}.
The LR-based DFT+U calculations for electronic and magnetic structures are presented 
and compared with previous theories and experiments in Sec.~\ref{subsec:el-mag}.
Finally, in Sec.~\ref{sec:CMYS}, systematical results for the other ternary and quaternary compounds are discussed 
and promising materials for HM ferromagnets are proposed.

%-- model & method ---------------------------------------------------------------------------------------
\section{model and method}\label{sec:model-method}

The full Heusler Co$_2$MnSi compound in $L2_1$ structure belongs to $Fm\bar{3}m$ ($O_h$) symmetry (space group 
No.~225).
For the modeling, a fcc-primitive cell that contains two Co atoms sitting at Wyckoff position ($1/4, 1/4, 1/4$) 
(multiplicity with Wyckoff letter is $8c$), one Mn atom at ($1/2, 1/2, 1/2$) ($4b$), and one $sp$-element Si at 
($0, 0, 0$) ($4a$) was prepared
(a conventional unit cell is shown in Fig.~\ref{fig_model}~(a)).
The detailed crystal structures for the ternary system Co$_2Y$Si, where Mn is replaced with
$Y$ of Ti, V, Cr, or Fe, and quaternary Co$_2$($Y$,Mn)Si, where a part of Mn is substituted with $Y$,
are described in Sec.~\ref{sec:CMYS}.

The self-consistent DFT calculations were performed via the {\it ab-initio}
package of Quantum-ESPRESSO\cite{qe-package} by implementing the ultra-soft pseudopotentials 
parametrized by the scheme of Rappe, Rabe, Kaxiras, and Joannopoulos~\cite{prb-41-1227R-1990,note_pp}.
The plane wave basis sets for the wave function and charge density had cutoff energies of 40~and 400~Ry,
respectively.
The self-consistent procedures were achieved until the iterative total energy difference became less than
the convergence criterion of $10^{-8}$~Ry, by using Monkhorst-Pack special $\mathbf{k}$-point 
mesh~\cite{prb-7-5212-1973} of $16\times16\times16$ in the first Brillouin zone by  
Methfessel-Paxton~\cite{prb-40-3616-1989} smearing method with a broadening parameter of 0.02~Ry.

The GGA functional formulated by Perdew, Burke, and Ernzerhof~\cite{prl-77-3865-1996}
was used for the exchange-correlation term. 
For the DFT+U method, a choice of ''double-counting'' correction term is also crucial to subtract the 
electron Coulomb energy that is already included in the LSDA or GGA functional.
This correction is conceptually desired to be the same energy contribution as that defined in LSDA or GGA. 
So far, however, an appropriate prescription for the double-counting term has not been established, but the so-called 
fully localized limit (FLL)~\cite{prb-52-R5467-1995,prb-48-16929-1993,prb-49-14211-1994,prb-50-16861-1994},
which is also referred as the atomic limit (AL), and 
around mean-field (AMF)~\cite{prb-44-943-1991,prb-49-14211-1994} approaches are mostly used.
The former functional favors the integer electron occupation numbers at a localized site, and thus, might be useful
for strongly correlated materials such as insulating oxide systems.
The latter might be for an intermediate of strongly correlated and itinerant materials. 
It is still under debate that which of two functionals is a proper approach for Heusler 
compounds.~\cite{jmmm-393-297-2015}
In this study, the double-counting functional incorporated in the simplified rotationally invariant
form~\cite{prb-57-1505-1998,JPhysCondMater-9-767-1997,prb-71-035105-2005}, which is equivalent to the FLL
approach but $J = 0$ (or approximately $U_\mathrm{eff}=U-J$, where $U_\mathrm{eff}$ stands for 
effective on-site Coulomb interaction), was employed.
We expect that this approach can easily address the underlying physics of correlated electronic structures, compared 
to AMF, because the electron-localization limit in FLL corresponds to the concept of Hubbard model;
thus, the scaling of $U_\mathrm{eff}$ can be simply understood as the strength of electron correlation.
The $U_\mathrm{eff}$ is computed within
the LR theory~\cite{prb-71-035105-2005,prb-97-035117-2018} for all transition metal atoms, where 
we assume that the Coulomb interaction is more dominant than the exchange at localized electron sites.

In the framework of LR theory~\cite{prb-71-035105-2005}, 
the on-site parameter for an atom $\alpha$, $U_\mathrm{eff}^\mathrm{LR(\alpha)}$, is evaluated from the second 
derivatives of the total energy functionals as
\begin{equation}
  U_\mathrm{eff}^\mathrm{LR (\alpha)}
  =  \frac{\partial^2 E^\mathrm{SCF}[\{q_\alpha\}]}{\partial q_\alpha^2}
    -\frac{\partial^2 E^\mathrm{KS}[\{q_\alpha\}]}{\partial q_\alpha^2}.
  \label{eq1}
\end{equation}
The total energies $E^\mathrm{SCF}$ and $E^\mathrm{KS}$ correspond to interacting (fully screened) and 
non-interacting systems.
The second term in Eq.~(\ref{eq1}) is necessary to subtract unphysical contributions in the total 
energy~\cite{prl-49-1691-1982,prl-51-1884-1983,prl-51-1888-1983}, which are caused by the conventional exchange-correlation 
functionals (LSDA and GGA), where the total energy has a curvature for non-integer occupation $q_\alpha$ and 
often misleads to incorrect energy minima.
The total energy derivatives are calculated using the constrained DFT approach:
\begin{equation}
  E^i[\lbrace q_\alpha \rbrace]
  = \min_{n(\mathbf{r}),\mu_\alpha} 
    \left\lbrace 
       E^i_\mathrm{GGA}[n(\mathbf{r})] + \sum_\alpha \mu_\alpha (n_\alpha - q_\alpha) 
    \right\rbrace,
  \label{eq2}
\end{equation}
where 
\begin{equation}
  \frac{\partial}{\partial q_\alpha} E^i [ \{ q_\alpha \} ] = -\mu_\alpha,
  \quad
  \frac{\partial^2}{\partial q_\alpha^2} E^i [ \{ q_\alpha \} ] 
  = -\frac{\partial \mu_\alpha}{\partial q_\alpha}.
  \label{eq3}
\end{equation}
The Lagrange multiplier $\mu_\alpha$ is a local perturbation potential that constrains the 
occupations $n_\alpha$ ($i=\mathrm{SCF, KS}$).
In practice, Eq.~(\ref{eq2}) is transformed into a tractable representation where the constraint fields are 
treated as independent variables by Legendre transformation and the variations of $n_\alpha$ with respect to
$\mu_\alpha$ are evaluated.~\cite{prb-71-035105-2005}
Using nonlocal linear response matrices
\begin{equation}
  (\chi_\mathrm{SCF})_{\beta \alpha} = \frac{\partial n_\beta}{\partial \mu_\alpha},
   \quad
  (\chi_\mathrm{KS})_{\beta \alpha}  = \frac{\partial n^\mathrm{KS}_\beta}{\partial \mu_\alpha},
  \label{eq4}
\end{equation}
Eq.~(\ref{eq1}) is rewritten to obtain $U_\mathrm{eff}^\mathrm{LR(\alpha)}$ as 
\begin{equation}
  U_\mathrm{eff}^\mathrm{LR (\alpha)}
  =  \left( \chi_\mathrm{KS}^{-1} - \chi_\mathrm{SCF}^{-1} \right)_{\alpha \alpha}.
  \label{eq5}
\end{equation}
The matrix elements of the response matrices are numerically computed;
$\chi_\mathrm{SCF}$ is obtained from the self-consistent (interacting) calculations under the applied local 
potential $\mu_\alpha$ and $\chi_\mathrm{KS}$ is obtained from the first iteration in a self-consistent cycle after
the end of GGA ground-state calculations -- the latter is occupation changes that arise from 
noninteracting hybridization due to $\mu_\alpha$.
The LR approach, in principle, requires a response of electron occupations with regard to the 
perturbed potentials at a single site in an infinite crystal environment for an accurate
$U_\mathrm{eff}$ evaluation, where all artifacts due to the periodic boundary conditions are 
excluded.~\cite{prb-71-035105-2005,prb-97-035117-2018}

%-------------------------------------------------------------------------------------------------------%
%   RESULTS AND DISCUSSIONS                                                                             %
%-------------------------------------------------------------------------------------------------------%

%-- LR-calculation for effective on-site Coulomb interaction --------------------------------------------
\section{Revisit of C\MakeLowercase{o}$_2$M\MakeLowercase{n}S\MakeLowercase{i}}\label{sec:CMS}

%-- LR-calculation for effective on-site Coulomb interaction --------------------------------------------
\subsection{LR-calculation for effective on-site Coulomb interaction parameter}\label{subsec:ulrt}

%... Fig.2
\begin{figure} [b]
\begin{center}
\includegraphics[width=1.0\columnwidth]{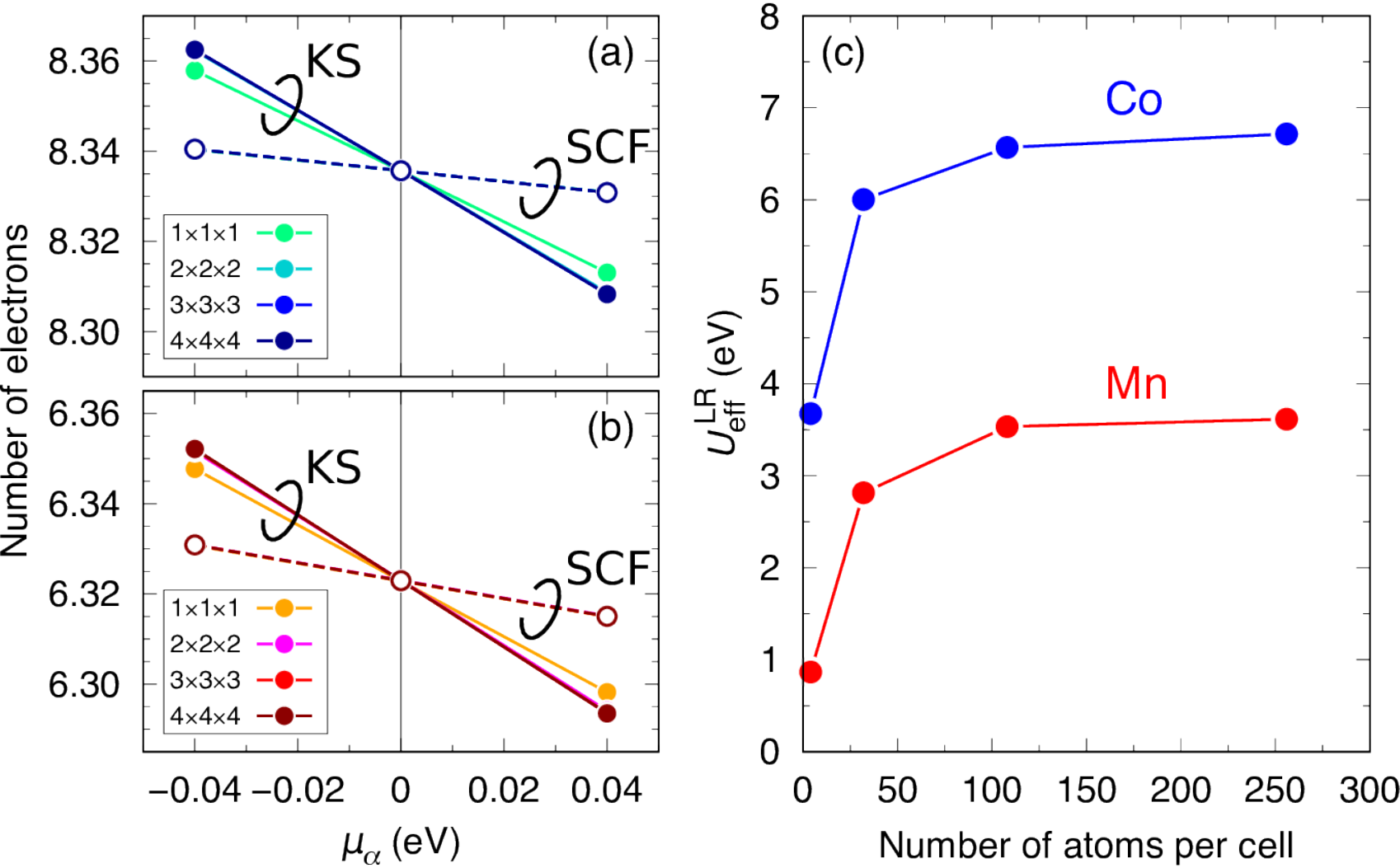}
\end{center}
\caption{(Color online) 
         Variations of occupied $3d$ electron numbers at (a)~on-site of Co and (b)~of Mn  
         as a function of applied perturbation potential $\mu_\alpha$ in $n\times n\times n$ 
         supercell Co$_2$MnSi ($n=1,2,3,4$).
         Solid lines indicate the KS calculation terms and dashed ones indicate SCF terms.
         (c)~The $U_\mathrm{eff}^\mathrm{LR}$ value dependence on the number of atoms per cell for Co (blue)
         and Mn (red).}
\label{sfig_ulrt}
\end{figure}

The LR calculations are performed to determine the correlation $U_\mathrm{eff}^\mathrm{LR(\alpha)}$ 
parameters.
The response functions of Eq.~(\ref{eq4}) are evaluated numerically from the gradient of $3d$ electron
occupation numbers with regard to the perturbed potential $\mu_\alpha$, which constrains the electrons of Co or Mn in Co$_2$MnSi alloy.
For the LR calculations, the experimental lattice constant $a_\mathrm{Expt.}=5.645$~{\AA}\cite{apl-88-032503-2006}
is employed.
As presented in Figs.~\ref{sfig_ulrt}~(a) and (b), the occupations' variation of the KS term in 
$1\times1\times1$ primitive cell is slightly off the others ($n\times n\times n$ cell where $n=2$, 3, 4),
although the SCF term does not change much (the plotted data are overlapping 
and the variations for different supercell sizes may not be visible from these figures).
The calculated $U_\mathrm{eff}^\mathrm{LR}$ value is plotted as a function of number of atoms per cell in 
Fig.~\ref{sfig_ulrt}~(c).
We find a $3\times3\times3$ fcc supercell including 108 atoms is practically large enough to obtain the 
well-converged parameters, meaning that the environment of the infinite crystal structure is well-reproduced.
The obtained values result in $U_\mathrm{eff}^\mathrm{LR(Mn)}=3.535$ and $U_\mathrm{eff}^\mathrm{LR(Co)}=6.570$~eV 
for Mn and Co, respectively.

The correlation parameter for Co is unexpectedly higher than the typically used empirical values, 
for instance,
the $U_\mathrm{eff}(=U-J)$ of 2.5~eV in full-potential (FP)-LMTO~\cite{jmmm-422-13-2017} and that of 2.1~eV  
in KKR~\cite{jphysDapplphys-42-084002-2009} calculations.
The constrained random phase approximation (cRPA) approach~\cite{prb-88-134402-2013} determines a 
parameter value similar to that determined in our study for Mn (3.07~eV), 
but almost half value of our LR result for Co (3.28~eV).
Table~\ref{table4} summarizes the numerical data of the $d$ occupations ($n_\alpha$) and the changes ($\Delta n_\alpha$)
induced by $\mu_\alpha$ in the LR calculations for $\alpha$ atom ($\alpha=$ Mn and Co).
As defined in Eqs.~(\ref{eq4}) and (\ref{eq5}), the $U_\mathrm{eff}^\mathrm{LR}$ is difference of the inversions of 
electron occupations' responses with respect to the applied potential shift $\mu_\alpha$ between KS and SCF terms.
For both $\alpha=$ Mn and Co cases, the absolute values of $\Delta n_\alpha$ in SCF are smaller than those in KS
by one order of magnitude,
so the inverted response function ($\chi^{-1} \propto \frac{1}{\Delta n_\alpha}$) of SCF becomes
a main factor in the computed correlation parameters.
We also find that, for the SCF term, $\Delta n_\mathrm{Co}$ is small compared to $\Delta n_\mathrm{Mn}$.
Therefore, the unreasonably overestimated parameter for Co originates from the difference in $\Delta n_\mathrm{Co}$ 
of the SCF term.
Using the diagonal matrix elements of $\chi^{-1}_\mathrm{KS}$ and $\chi^{-1}_\mathrm{SCF}$ in Eq.~(\ref{eq5}),
the parameter for Co is calculated as $U_\mathrm{eff}^\mathrm{LR(Co)}=-1.10114-(-7.67152)=6.570$~eV.
For Mn, $U_\mathrm{eff}^\mathrm{LR(Mn)}=-0.98733-(-4.52196)=3.535$~eV is obtained, where the inverted KS response 
function's contribution (the first term) is almost the same as the $U_\mathrm{eff}^\mathrm{LR(Co)}$ case, 
while the SCF one (the second term) is significantly different.

%... TABLE 1
\begin{table} 
\caption{Numerical data of LR calculations for $U_\mathrm{eff}^{\mathrm{LR(}\alpha\mathrm{)}}$ 
         parameters ($\alpha=$ Mn or Co);
         $d$ occupation numbers ($n_\alpha$) and changes ($\Delta n_\alpha$) of the {\it on-site} 
         $\alpha$ atom from neutral state ($\mu_\alpha=0$~eV) of KS and SCF terms 
         when the perturbed potential is applied to the {\it on-site}
         $\alpha$ atom ($\mu_\alpha \ne 0$~eV). 
         The results are obtained from the $3\times 3\times 3$ supercell,
         in which the well-converged parameters are computed.}
\begin{threeparttable}
\begin{ruledtabular}
\renewcommand{\arraystretch}{1.3}
\begin{tabular}{lrrrrr}
 & \multirow{2}{*}{$\mu_\alpha~$(eV)} & \multicolumn{2}{c}{KS} & \multicolumn{2}{c}{SCF} \\
\cline{3-4} \cline{5-6}
 & & \multicolumn{1}{c}{~$n_\alpha$} & \multicolumn{1}{c}{~$\Delta n_\alpha$}   
   & \multicolumn{1}{c}{~$n_\alpha$} & \multicolumn{1}{c}{~$\Delta n_\alpha$}   \\
\hline
Mn &  $-0.04$  &  6.35202  &  $ 0.02908$  &  6.33086  &  $ 0.00792$  \\
   &  $ 0.00$  &  6.32294  &  $ 0.00000$  &  6.32294  &  $ 0.00000$  \\
   &  $ 0.04$  &  6.29357  &  $-0.02937$  &  6.31500  &  $-0.00794$  \\
\hline
Co &  $-0.04$  &  8.36249  &  $ 0.02682$  &  8.34047  &  $ 0.00480$  \\
   &  $ 0.00$  &  8.33567  &  $ 0.00000$  &  8.33567  &  $ 0.00000$  \\
   &  $ 0.04$  &  8.30830  &  $-0.02737$  &  8.33085  &  $-0.00482$  \\
\end{tabular}
\end{ruledtabular}
\end{threeparttable}
\label{table4}
\end{table}

From above discussion, we conclude that the overestimation of $U_\mathrm{eff}^\mathrm{LR}$ for the Co site arises 
from the fact that the charge density response of Co is insensitive compared to Mn or is still insufficient to
evaluate the parameters through the SCF iteration cycles under the applied potential  
shift.~\cite{note_ulrt_alpha-size}
This can be attributed to the delocalized electronic structures of Co compared to Mn, which originates from that 
the Co $d$ orbital distribution is spatially-spread due to the $d$ orbital hybridization with first (Mn)
and second (Co) neighboring atoms, whereas the Mn $d$ orbital distribution is spatially-narrow due to the 
$d$ hybridization with 
only first (Co) neighboring atom, as discussed in Sec.~\ref{subsec:org-HM}.
The localized characters of Co electronic states compared to those of Mn are consistent with the fact that
the spin magnetic moment of Co ($1.05~\mu_\mathrm{B}$) is much smaller than that of Mn ($2.95~\mu_\mathrm{B}$).
Recently, an extended LR theory~\cite{pccp-19-8008-2017} has been proposed to overcome the insufficiency
of response of charge density;
the second response of charge density is additionally included, which is required for complete cancelation 
of the electron-electron Coulomb interaction (Hartree energy) term changed by the external potential 
($\mu_\alpha \ne 0$), which might be canceled incompletely in the current LR calculation.
In the present study, the DFT+U method incorporated by FLL formalism is used, 
but another approach for solving the failure in estimating $U_\mathrm{eff}^\mathrm{LR}$ for a weakly correlated Co site
might be the use of AMF approximation\cite{prb-44-943-1991,prb-49-14211-1994}
or {\it hybrid} approach~\cite{prb-67-153106-2003}, where the AMF and FLL approximations are linearly interpolated.

%-- Structural property ---------------------------------------------------------------------------------
\subsection{Structural property}\label{subsec:str}

%... Fig.3
\begin{figure} [b]
\begin{center}
\includegraphics[width=0.85\columnwidth]{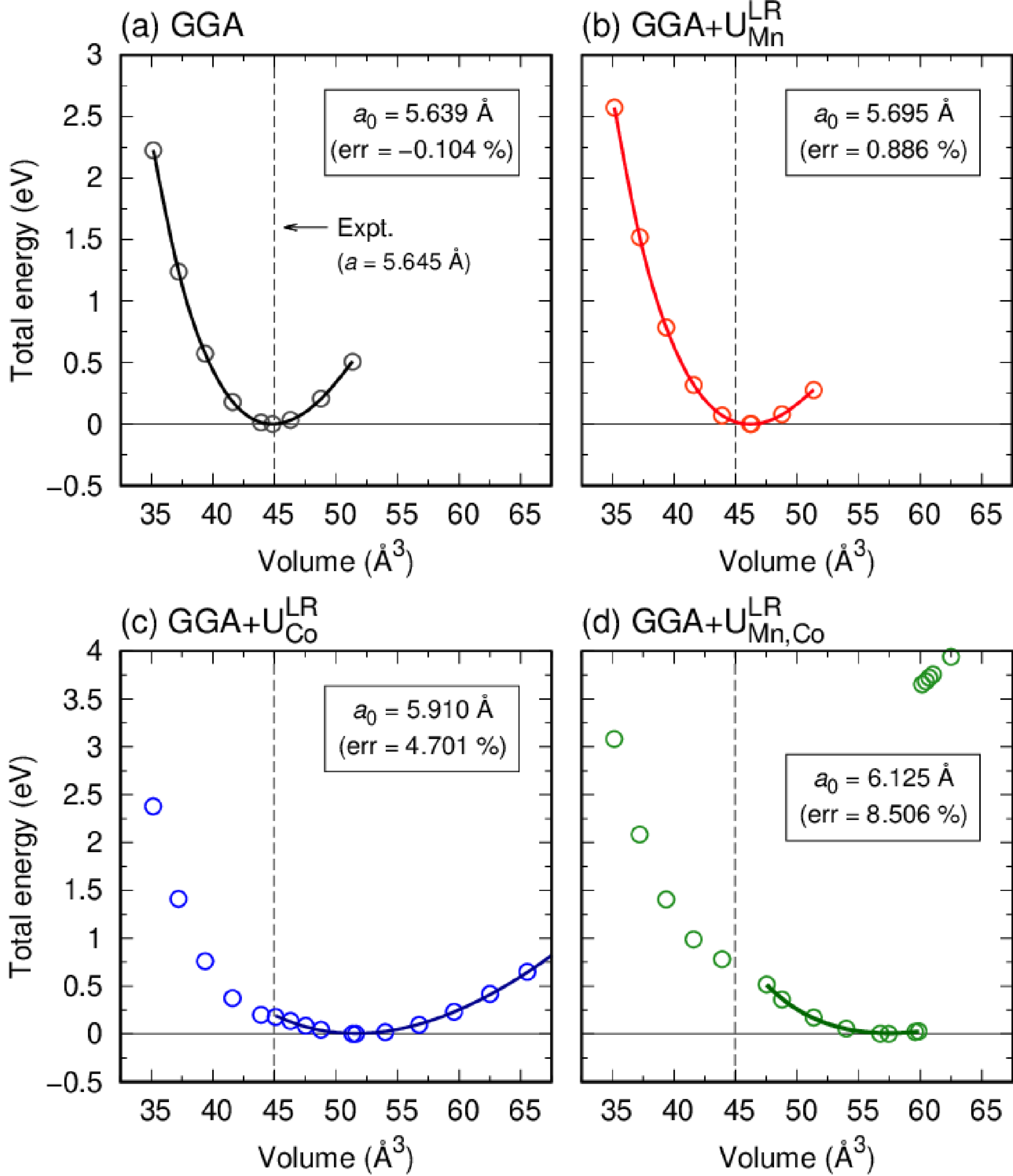}
\end{center}
\caption{(Color online) 
         Total energy as a function of volume for Co$_2$MnSi from 
         (a)~GGA,
         (b)~GGA+U$_\mathrm{Mn}^\mathrm{LR}$
         (c)~GGA+U$_\mathrm{Co}^\mathrm{LR}$, and
         (d)~GGA+U$_\mathrm{Mn,Co}^\mathrm{LR}$.
         Opened circles are obtained from the first principles and solid lines are from Murnaghan
         fitting, which determines the equilibrium lattice constant $a_0$, as shown in the inset.
         The error from the experiment is also shown in the parentheses.
         The experimental value is plotted by a dashed (black) line.}
\label{sfig_alat}
\end{figure}

%... TABLE 2
\begin{table*}
\caption{Structural parameters of lattice constant $a_0$, bulk modulus $B_0$, and 
         its pressure derivative $B_0'$ for Co$_2$MnSi.
         The $U_\mathrm{eff}^\mathrm{LR}$ values determined by the LR approach
         are employed in the present study: 
         $U_\mathrm{eff}^\mathrm{LR(Mn)}=3.535$~eV for GGA+U$_\mathrm{Mn}^\mathrm{LR}$,
         $U_\mathrm{eff}^\mathrm{LR(Co)}=6.570$~eV for GGA+U$_\mathrm{Co}^\mathrm{LR}$, and
         both for GGA+U$_\mathrm{Mn,Co}^\mathrm{LR}$.}
\begin{threeparttable}
\begin{ruledtabular}
\renewcommand{\arraystretch}{1.3}
\begin{tabular}{llccc}
   & & $a_0$~(\AA) & $B_0$~(GPa) & $B_0'$ \\
\hline
Present work  & GGA                                & 5.639 & 217.63 & 4.30  \\
              & GGA+U$_\mathrm{Mn}^\mathrm{LR}$    & 5.659 & 186.29 & 4.41  \\
              & GGA+U$_\mathrm{Co}^\mathrm{LR}$    & 5.910 &  65.40 & 1.25  \\
              & GGA+U$_\mathrm{Mn,Co}^\mathrm{LR}$ & 6.125 &  60.70 & 5.74  \\
\hline
 Theory  & LSDA & 5.54\tnote{a}  & 258.0\tnote{a}  &        \\
         &  GGA & 5.643\tnote{b} , 5.633\tnote{c} , 5.639\tnote{d} ,  5.642\tnote{e}  
                &  226\tnote{b}  , 212.8\tnote{c} , 214\tnote{d}   , 240.89\tnote{e} 
                &                  4.680\tnote{c} , 4.674\tnote{d} ,  4.983\tnote{e} \\
\end{tabular}
\begin{tablenotes}[flushleft,para]
  \item[a] Reference~[\onlinecite{IEEETransMagn-50-1301104-2014}]
  \item[b] Reference~[\onlinecite{prb-66-094421-2002}]
  \item[c] Reference~[\onlinecite{JAlloyCompd-560-25-2013}]
  \item[d] Reference~[\onlinecite{EuroPhysJB-76-321-2010}]
  \item[e] Reference~[\onlinecite{intermetal-44-26-2014}]
\end{tablenotes}
\end{ruledtabular}
\end{threeparttable}
\label{table:str_CMS}
\end{table*}

Here, we consider three schemes of LR-based DFT+U calculations, in addition to the standard GGA:  
the determined $U_\mathrm{eff}^\mathrm{LR}$ values are applied 
to only Mn (referred as GGA+U$_\mathrm{Mn}^\mathrm{LR}$) or Co (GGA+U$_\mathrm{Co}^\mathrm{LR}$) and 
to both of them (GGA+U$_\mathrm{Mn,Co}^\mathrm{LR}$).
First, the GGA calculations are performed for evaluating the equilibrium lattice constant.
The total energies at different volume sizes of a primitive cell are obtained as shown in Fig.~\ref{sfig_alat}~(a).
The energy minimum is searched by energy fitting to the Murnaghan equation of
states~\cite{pnasusa-30-244-1944} as a function of volume $V$, 
\begin{equation}
  E(V)=E_0+\frac{B_0V}{B_0'}\left[ \frac{1}{B_0'-1}\left(\frac{V_0}{V}\right)^{B_0'}+1 \right]-\frac{B_0}{B_0'-1}V_0,
\label{eq_Murnaghan}
\end{equation}
where $E_0$ is the ground-state total energy at equilibrium volume $V_0$, $B_0$ bulk modulus, and $B_0'$ 
pressure derivative of the bulk modulus.
The obtained lattice constant is 5.639~{\AA}, which agrees with the experimental value.~\cite{apl-88-032503-2006}
The error value between the calculated lattice constant $a_0$ and experiment, defined as 
$(a_0-a_\mathrm{Expt.})/a_\mathrm{Expt.}\times 100~(\%)$, 
is only $-0.104~\%$.
In the GGA+U$_\mathrm{eff}^\mathrm{LR}$ case, the obtained lattice constant of 5.695~{\AA} is similar to the GGA 
result and the error from the experiment is less than 1~\% (0.866~\%), as shown in Fig.~\ref{sfig_alat}~(b).
On the other hand, $a_0$ is significantly overestimated by the errors of
4.701 and 8.506~\% in the GGA+U$_\mathrm{Co}^\mathrm{LR}$ and GGA+U$_\mathrm{Mn,Co}^\mathrm{LR}$ cases.
Figure~\ref{sfig_alat}~(c) shows a local energy minimum around the experimental value but the global minimum
is found at 51.62~{\AA}$^3$, corresponding to $a_0=5.910~\mathrm{\AA}$.
Note also that in the GGA+U$_\mathrm{Mn,Co}^\mathrm{LR}$ scheme (Fig.~\ref{sfig_alat}~(d)), a jump of total energy 
change around the volume of $\sim60~\mathrm{\AA}^3$ occurs due to a magnetic phase transition, but we confirm that an energy 
minimum, corresponding to $a_0=6.125$~{\AA}, exists at less than the volume where this magnetic transition is induced.

We present bulk modulus $B_0$ and its pressure derivative $B_0'$ in Table~\ref{table:str_CMS}, through comparisons of 
theoretical literature.~\cite{prb-66-094421-2002,IEEETransMagn-50-1301104-2014,JAlloyCompd-560-25-2013,
EuroPhysJB-76-321-2010,intermetal-44-26-2014}
Among the previous reports, the $B_0$ of LSDA is greater than that of GGA.
Our GGA result is almost similar to the reported values in $B_0$ and $B_0'$,
while the GGA+U$_\mathrm{Mn}^\mathrm{LR}$ result is slightly smaller in $B_0$.
On the other hand, the $B_0$ calculated by GGA+U$_\mathrm{Co}^\mathrm{LR}$ and GGA+U$_\mathrm{Mn,Co}^\mathrm{LR}$ 
methods is one order of magnitude smaller than the other calculations.
Because the experimental data of the bulk modulus and its derivative are not available for Co$_2$MnSi at this moment,
we cannot conclude the validity of our method.
However, at least focusing on the lattice constant, 
these results indicate that the introduction of $U_\mathrm{eff}^\mathrm{LR}$
to the Mn atom tends to obtain a reasonable result, as well as GGA, from the comparison of experiments,
while the inclusion of $U_\mathrm{eff}^\mathrm{LR}$ to Co fails to evaluate the $a_0$ of Co$_2$MnSi.

%-- Origin of half-metallicity --------------------------------------------------------------------------
\subsection{Origin of half-metallicity}\label{subsec:org-HM}

%... Fig.4
\begin{figure*}
\includegraphics[width=1.9\columnwidth]{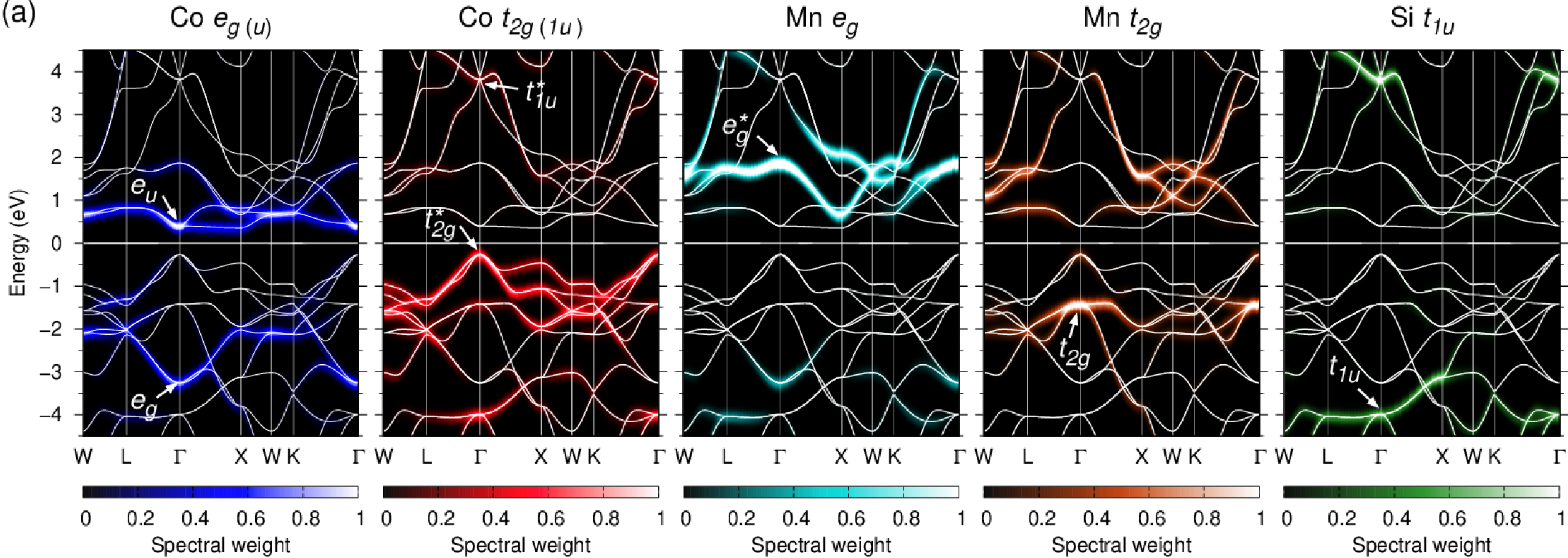}\\~\\
\includegraphics[width=1.8\columnwidth]{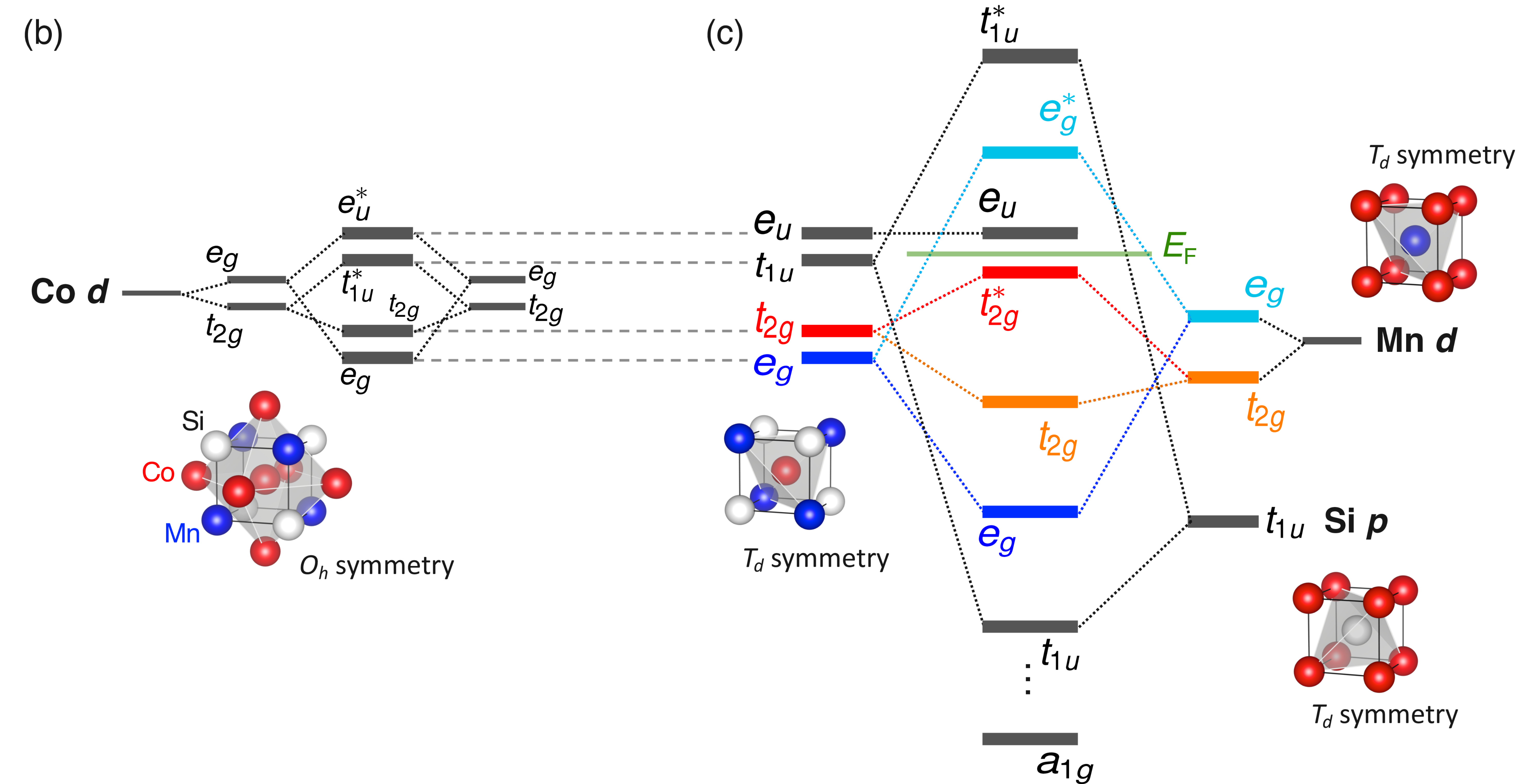}
\caption{(Color online)
         (a)~Projected band structures for minority spin in Co$_2$MnSi.
         The orbital-component spectral weights of $e_{g(u)}$ (blue) and 
         $t_{2g(1u)}$ (red) symmetries for Co $d$ orbitals, $e_g$ (skyblue) and $t_{2g}$ (orange) for Mn $d$, 
         and $t_{1u}$ (green) for Si $p$ are shown by the colormap.
         Total band structure of minority spin is also plotted by a white solid line.
         The Fermi energy is set to zero.
         Minority-spin-state atomic-orbital energy diagrams of 
         (b)~hybridizations of $d$ orbitals between two Co atoms at different sublattices in $O_h$ site symmetry
         and 
         (c)~hybridizations among Co--Co~$d$, Mn~$d$, and Si~$p$ in $T_d$ site symmetry, where $a_{1g}$ 
         corresponds to Si $s$ orbital, which does not appear in the projected bands given in (a).
         Note the orbital symmetry characters are represented under the $O_h$ site symmetry 
         throughout the diagram:
         Representations of $d$--$e$ ($d_{z^2}, d_{x^2-y^2}$), --$t_2$ ($d_{xz}, d_{yz}, d_{xy}$), and 
         $p$--$t_2$ ($p_x, p_y, p_z$) states in $T_d$ site symmetry can be transformed into those of $e_g$, $t_{2g}$,
         and $t_{1u}$ states in $O_h$ site symmetry, respectively.
         The asterisks indicating the anti-bonding state in (b) are omitted in (c) for simplicity.}
\label{fig_band_gga}
\end{figure*}

%... TABLE 1
\begin{table}
\caption{Total and atom-resolved magnetic moments (in $\mu_\mathrm{B}$) for Co$_2$MnSi 
         with the comparison of present and previous theories as well as experiments.
         The representations in the present paper are the same as those in Table~\ref{table:str_CMS}.
         The first column gives the calculation methods (and types of exchange-correlation functionals
         in parentheses) for theory and measurement techniques for experiment.}
\begin{threeparttable}
\begin{ruledtabular}
\renewcommand{\arraystretch}{1.3}
\begin{tabular}{lccccc}
  & Total & Co & Mn & Si & Ref. \\
\hline
Present work & & & & &\\
  GGA                                & 5.01 & 1.05 & 2.95 & $-0.05$ & \\ 
  GGA+U$_\mathrm{Mn}^\mathrm{LR}$    & 5.01 & 0.72 & 3.63 & $-0.08$ & \\ 
  GGA+U$_\mathrm{Co}^\mathrm{LR}$    & 6.95 & 1.88 & 3.19 & $-0.09$ & \\ 
  GGA+U$_\mathrm{Mn,Co}^\mathrm{LR}$ & 8.08 & 1.94 & 4.05 & $-0.06$ & \\ 
\hline
Theory\tnote{a} & & & & &\\
  FS-KKR (LSDA)                & 4.94 & 1.02 & 2.97 & $-0.07$ & [\onlinecite{prb-66-174429-2002}] \\
  ASA-ASW (GGA)                & 5.00 & 0.93 & 3.21 & $-0.06$ & [\onlinecite{prb-72-184415-2005}]  \\
  FLAPW (GGA)                  & 5.00 & 1.06 & 2.92 & $-0.04$ & [\onlinecite{prb-66-094421-2002}]  \\
  FP-LMTO (GGA+U\tnote{b} )    & 5.00 & 1.08 & 2.97 & $-0.08$ & [\onlinecite{jmmm-422-13-2017}] \\
  MLWF-FLAPW (GGA+U\tnote{c} ) & 5.00 & 1.05 & 3.01 & $-0.06$ & [\onlinecite{prb-88-134402-2013}] \\
  KKR (LSDA+DMFT\tnote{d} )    & 4.97 &      &      &         & [\onlinecite{jphysDapplphys-42-084002-2009}] \\
  FLAPW-GW (GGA)               & 5.00 &      &      &         & [\onlinecite{prb-86-245115-2012}] \\
\hline
Experiment\tnote{e} & & & & &\\
  Sucksmith & 5.07 & 0.75 & 3.57 & & [\onlinecite{jpcs-32-1221-1971}] \\
            & 5.01 &      &      & & [\onlinecite{jmmm-937-54-1986}] \\
  SQUID     & 4.97 &      &      & & [\onlinecite{prb-74-104405-2006}] \\
  SQUID     & 5.00 & 0.72 & 3.34 & & [\onlinecite{prb-80-144405-2009}] \\
\end{tabular}
\begin{tablenotes}
\item[a]
  FS-KKR: Full-potential screened Korringa-Kohn-Rostoker Green's function method;
  ASA: atomic sphere approximation;
  ASW: augmented spherical waves method; 
  FLAPW: full-potential linearized augmented plane wave method;
  FP-LMTO: Full-potential liner muffin-tin orbital method;
  MLWF: maximally localized Wannier functions;
  GW: GW approximation.
\item[b] 
  The $U$ and $J$ values of 3.5 (5.0) and 1.0 (0.9)~eV for Co (Mn), respectively,
  are chosen to reproduce the total spin magnetic moment observed experimentally.
\item[c] 
  The respective $U_\mathrm{eff}$ values of 3.28 and 3.07~eV for Co and Mn are determined by 
  cRPA.
\item[d] 
  The $U$ and $J$ values of 3.0 and 0.9~eV, which have been reported as average values of 
  The determined parameters by theory for $3d$ transition metal pure bulks, are used.
\item[e] 
  Sucksmith: Sucksmith ring-balance measurement by Faraday method;
  SQUID: Superconducting quantum interface device magnetometry.
\end{tablenotes}
\end{ruledtabular}
\end{threeparttable}
\label{table1}
\end{table}

As mentioned in Sec.~\ref{sec:model-method}, the full Heusler alloy of $L2_1$ structure belongs to the 
octahedral ($O_h$) space group symmetry.
In this {\it whole-crystal} symmetry, 
we first focus on the Co lattice by ignoring the first-neighboring Mn and Si atoms.
The lattice is assumed to be a simple cubic composed by the second-neighboring Co at different sublattices in 
the primitive cell, which leads to the Co sitting at $O_h$ {\it site} symmetry.
Second, our focus is turned on the tetrahedral ($T_d$) {\it site} symmetry.
Neglecting the chemical atomic species, every atom forms a bcc lattice structure and is surrounded by
a tetrahedral environment.
The hybridization diagram of atomic orbital energy is discussed by following these two steps.
Note that, for avoiding the confusion regarding the notations, the symmetric characters of the atomic orbital are 
unified using only representations for the $O_h$ site symmetry, which corresponds to the space group of the $L2_1$ 
full Heusler compound, as done also in the previous research~\cite{prb-66-174429-2002}.

We again start with the results of the standard GGA calculations for discussing the underlying electronic 
structure of Co$_2$MnSi.
Figure~\ref{fig_band_gga}~(a) shows the band structures for the minority spin states projected into 
the Co~$e_{g(u)}$ ($d_{z^2}, d_{x^2-y^2}$) and $t_{2g(1u)}$ ($d_{xz}, d_{yz}, d_{xy}$), 
Mn~$e_g$ ($d_{z^2}, d_{x^2-y^2}$) and $t_{2g}$ ($d_{xz}, d_{yz}, d_{xy}$), and 
Si~$t_{1u}$ ($p_x, p_y, p_z$) orbitals.
The lattice constant is set to the theoretically obtained value of 5.639~{\AA}.
The $d$ states of Co and Mn are visible around Fermi energy, while the Si $t_{1u}$ state can be seen only at very 
far from Fermi energy.
To discuss the orbital hybridization mechanism, the eigenstates at $\Gamma$ point in Brillouin zone are focused. 
At 0.4~eV above the Fermi energy, the Co $e_u$ state appears but the other orbital components are not included 
in these eigenstates,
which means that Co~$e_u$ does not hybridize with the other atomic orbitals.
We can find $t_{2g}$ hybridization between Co and Mn that forms a bonding Mn~$t_{2g}$ state at $-1.4$~eV and 
an anti-bonding Co~$t_{2g}^*$ at $-0.3$~eV.
As a result, a minority band gap is originated by the anti-bonding $t_{2g}^*$ and non-bonding $e_u$
states of Co atom.
Another essential orbital hybridization is found in the $t_{1u}$ symmetry character between Co and Si.
The eigenstate components of Co and Si exist at energy levels of 3.8 and $-3.9$~eV, respectively, 
so that the Co and Si atoms contribute to the anti-bonding state ($t_{1u}^*$) and to the bonding state ($t_{1u}$), respectively.
We do not mention the $e_g$ hybridization between Co and Mn as it has already been discussed previously.~\cite{prb-66-174429-2002}

Figure~\ref{fig_band_gga}~(b) presents an energy diagram of Co atoms under $O_h$ site symmetry.
Due to the crystal field, the $e_g$ and $t_{2g}$ orbitals are formed and hybridize with the same character
orbitals of Co at the other site. 
These hybridizations arise from the bonding states of $e_g$ and $t_{2g}$ orbitals and anti-bonding states of $e_u^*$ 
and $t_{1u}^*$ orbitals.
The $t_{2g}$ orbital hybridization, including $d_{xz}$--$d_{xz}$, $d_{yz}$--$d_{yz}$, and $d_{xy}$--$d_{xy}$,
is expected to form a $\pi$-like bonding in the $O_h$ atomic positions, 
and $e_g$ hybridization, including $d_{z^2}$--$d_{z^2}$ and $d_{x^2-y^2}$--$d_{x^2-y^2}$, 
is expected to form a $\sigma$-like bonding, whose orbital coupling is stronger than that of $\pi$-like bonding.
Accordingly, the energy gap between bonding $e_g$ and anti-bonding $e_u^*$ states arising from $e_g$ hybridization
becomes wide compared to the bonding $t_{2g}$ and anti-bonding $t_{1u}^*$ states from $t_{2g}$ hybridization.

Next, the orbital interactions between the first-neighboring atom are discussed by focusing on $T_d$ site symmetry.
Before that, we mention here the correlation between the $O_h$ and $T_d$ site symmetries and the possibility 
of atomic orbitals to hybridize.
The $T_d$ site symmetry, which is a subgroup of the $O_h$ site symmetry, has the same irreducible representations 
as the $O_h$ site symmetry except for an absence (presence) of inversion symmetry in $T_d$ ($O_h$) site symmetry.
The Co--Co~$d$ orbital's character in the $O_h$ site symmetry can be transformed into the $T_d$ notation;
the doublet $e_g$ and $e_u^*$ orbitals in $O_h$ are represented as the $e$ state in $T_d$, and 
the triplet $t_{2g}$ and $t_{1u}^*$ orbitals as $t_2$, where the asterisk symbol indicating the anti-bonding state
is omitted for simplicity.
The $T_d$ site symmetry also gives the $e$ ($d_{z^2}, d_{x^2-y^2}$) and $t_2$ ($d_{xz}, d_{yz}, d_{xy}$) 
characters for Mn and $t_2$ ($p_x, p_y, p_z$) for Si.
These augments allow Co--Mn and Co--Si to interact in the atomic orbitals in $T_d$ site symmetry,
i.e., $t_{2g}$ orbital hybridization of Co--Mn and $t_{1u}$ orbital hybridization of Co--Si in $O_h$ site symmetry.

Figure~\ref{fig_band_gga}~(c) illustrates the possible energy diagram between Co--Co and Mn or Si.
The $a_{1g}$ orbital corresponds to the Si~$s$ orbital, which does not appear 
in the band structure of Fig.~\ref{fig_band_gga}~(a) because the energy level is very low.
The anti-bonding Co~$t_{2g}^*$ state dominates the highest orbital state in the valence band, which hybridizes with bonding Mn~$t_{2g}$.
The Co~$t_{1u}^*$ is pushed up to a quite higher energy through the hybridization with Si~$t_{1u}$ ($p$) orbital
and the non-bonding Co~$e_u$ is left at above the Fermi level.
This energy diagram, thus, suggests that the main contributions to constructing the minority band gap arise from the $t_{2g}$ coupling of Co and Mn atoms and the $t_{1u}$ orbital of Co no longer contributes to the gap;
this conclusion is different from that of a previous study~\cite{prb-66-174429-2002}, where the band gap in the minority state 
is mostly dominated by Co $e_u$ and $t_{1u}$ orbitals.
Instead, more importantly, our diagram proposes that the HM property and electronic structure near the Fermi 
level can be tuned by a selection of $Y$ atom and/or a mix of several atoms into $Y$ site 
through $t_{2g}$ coupling in $L2_1$ Heusler alloy.

Even though our diagram is at variance with the previously reported one~\cite{prb-66-174429-2002}, 
the 12 valence electrons for Co$_2$MnSi are confirmed to occupy three Co $t_{2g}^*$, three Mn $t_{2g}$,
two Co $e_g$, three Si $t_{1u}$, and one Si $a_{1g}$ orbitals in the down-spin state.
This means our diagram satisfies the well-known Slater-Pauling relation~\cite{prb-66-174429-2002}:
The magnetic moment of the system, $m_\mathrm{spin}$, is obeyed by $m_\mathrm{spin}=N_\mathrm{val}-24$, 
where $N_\mathrm{val}$ is the total number of valence electrons.
The calculations obtain a total magnetic moment of 5.01~$\mu_\mathrm{B}$, which is very close to the integer 
value expected by the Slater-Pauling rule and in agreement with a previous theory within
LSDA~\cite{prb-66-174429-2002} and GGA~\cite{prb-72-184415-2005,prb-66-094421-2002}, as well as the  
experiments~\cite{prb-74-104405-2006,jmmm-937-54-1986,jpcs-32-1221-1971,prb-80-144405-2009}, 
as summarized in Table~\ref{table1}.

Note that the previous study~\cite{prb-66-174429-2002} was carried out for Co$_2$MnGe, where
the number of valence electrons is equivalent to that of Co$_2$MnSi.
Thus, Co$_2$MnGe is confirmed to be similar to Co$_2$MnSi.
The energy diagram obtained from the band structure calculations corresponds to Fig.~\ref{fig_band_gga}~(c)
and the integer value of the total spin magnetic moment is calculated ($m_\mathrm{spin}=5.00~\mu_\mathrm{B}$).

%-- Role of correlation effects on Mn and Co ------------------------------------------------------------
\subsection{Correlation effects on Mn and Co}\label{subsec:effect-of-U}

As mentioned in the introduction, the behaviors of electron localizations are supposed to be different at
Co and Mn sites.
This fact motivates us to investigate the effects of the correlations for each site.
To discuss the influence of +U on the atomic energy diagram, modifications of magnetic moment and
band structure
are studied by performing DFT+U calculations with varying $U_\mathrm{eff}$ parameters
for Co and Mn atoms {\it independently}.
Here, we refer to the case where the varying $U_\mathrm{eff}$ is applied to only Mn (Co) site as GGA+U$_\mathrm{Mn}$
(GGA+U$_\mathrm{Co}$) representation, 
where the lattice constant is set to the theoretical value of 5.695~{\AA} (5.910~{\AA}) obtained in 
Sec.~\ref{subsec:str}.

%... Fig.5
\begin{figure} [b]
\begin{center}
\includegraphics[width=0.9\columnwidth]{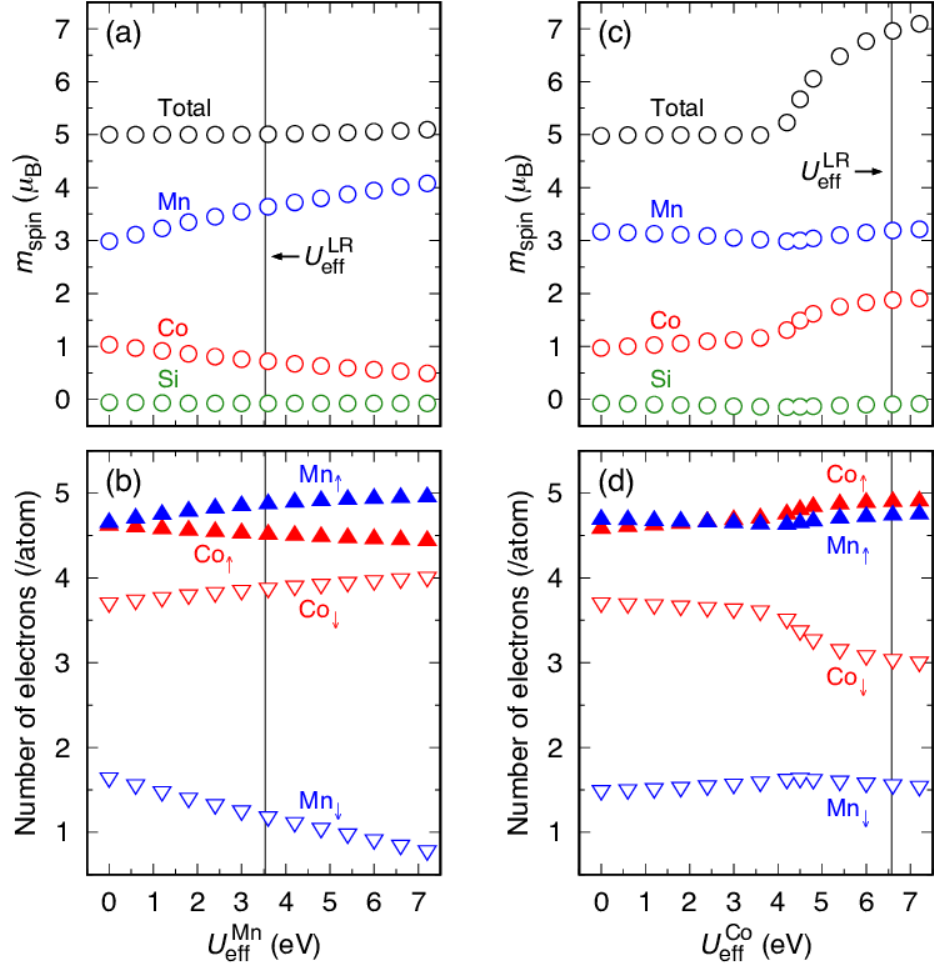}
\end{center}
\caption{(Color online)
         Dependence of (a)~total and atom-resolved spin magnetic moments, $m_\mathrm{Spin}$,
         and (b)~$d$ orbital occupations with respect to the varying $U_\mathrm{eff}^\mathrm{Mn}$
         for GGA+U$_\mathrm{Mn}$ case.
         Black, red, blue, and green circles in (a) indicate total, Co, Mn, and Si, and
         red and blue up-(down-)pointing triangles in (b) are majority (minority) $d$ occupations for Co and Mn, 
         respectively.
         The vertical solid line indicates the value of $U_\mathrm{eff}^\mathrm{LR}$.
         (c) and (d)~Same plots for GGA+U$_\mathrm{eff}^\mathrm{Co}$ having the same notations as those in (a) and (b).
         $U_\mathrm{eff}^\alpha=0$ ($\alpha$ is Mn or Co) indicates the GGA result, where the difference between
         GGA+U$_\mathrm{Mn}$ and GGA+U$_\mathrm{Co}$ comes from the different equilibrium lattice constants.}
\label{fig_mspin-ueff}
\end{figure}

%... Fig.4
\begin{figure*}
\begin{center}
\end{center}
\includegraphics[width=1.5\columnwidth]{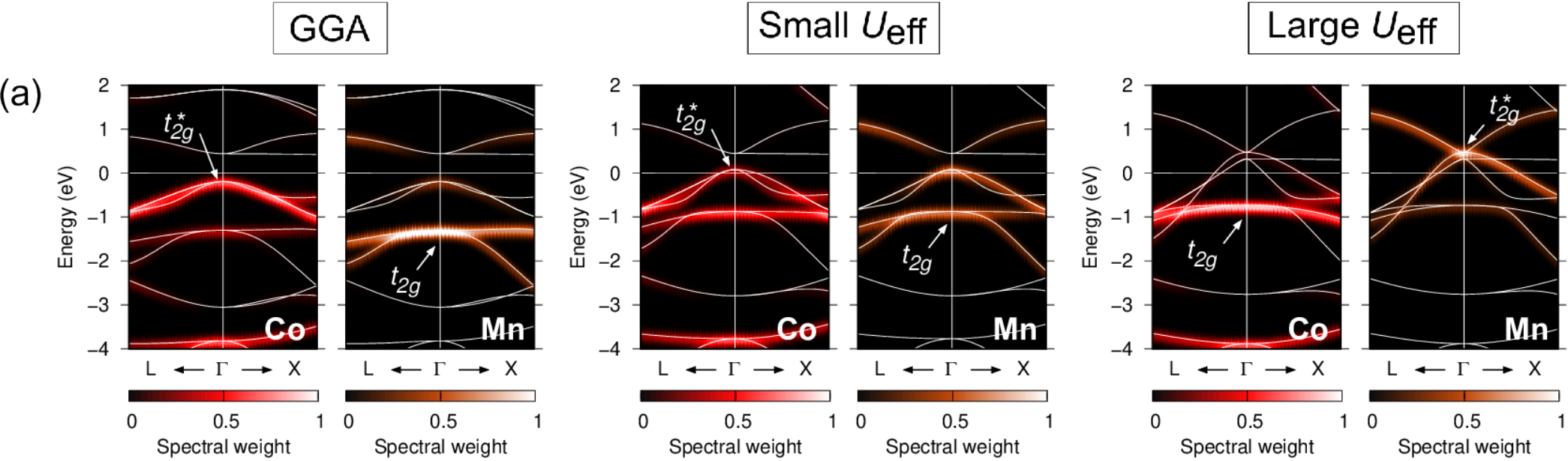}
\\~\\
\includegraphics[width=1.46\columnwidth]{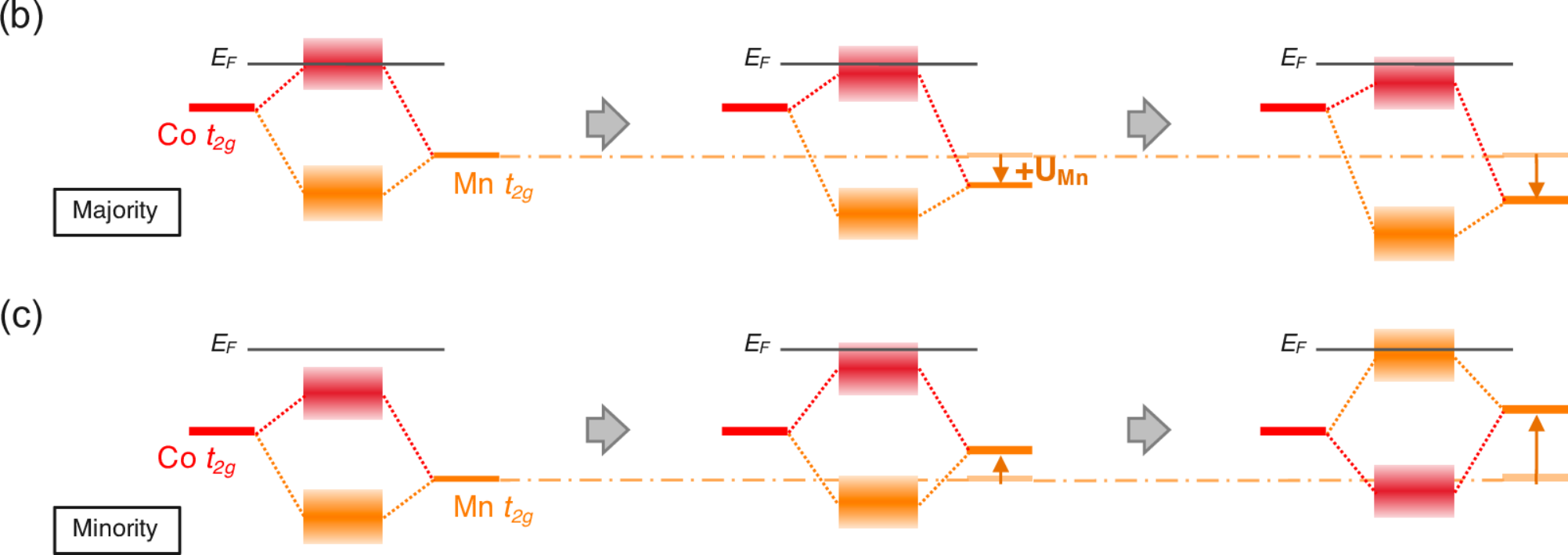}
\caption{(Color online) 
         (a)~Dependence of band structures in minority state of varying 
         +U$_\mathrm{Mn}$ parameter, i.e., GGA ($U_\mathrm{eff}=0$~eV for Mn site), small (3~eV) 
         and large (6~eV) values, where the projected spectral weights for Co and Mn $t_{2g}$ 
         states are shown in left (red) and right (orange) panels, respectively.
         The Fermi energy is set to zero and total minority band structure is plotted by a white line.
         Schematic summary of changes in atomic orbital hybridizations for (b)~majority and (c)~minority states.
         Arrows in (b) and (c) indicate the energy shift induced by the effect of +U$_\mathrm{Mn}$.}
\label{fig_diagram_Mn-U}
\end{figure*}

We first mention the GGA+U$_\mathrm{Mn}$ case.
The total $m_\mathrm{spin}$ is constant but Mn (Co) $m_\mathrm{spin}$ monotonically increases (decreases)
when the correlation parameter for Mn, $U_\mathrm{eff}^\mathrm{Mn}$, increases (see Fig.~\ref{fig_mspin-ueff}~(a)).
Note that two of the Co atoms exist in the primitive cell, so the variation of Co $m_\mathrm{spin}$ is estimated to be twice.
The increased $m_\mathrm{spin}$ of Mn arises from a significant reduction in minority
spin electron occupations, 
as shown in Fig.~\ref{fig_mspin-ueff}~(b).
This reflects the following behavior:
a large +U value intensifies the Coulomb interaction %aspect 
contributions and allows electrons to occupy not same but
different orbitals with parallel spins from Pauli exclusion principles and Hund's rule, 
leading to a gain in kinetic energy.

As the $t_{2g}$ orbitals of Co and Mn change the most noticeably %by different 
depending on the $U_\mathrm{eff}^\mathrm{Mn}$
value, we trace modifications in the band structures of these orbitals.
Figure~\ref{fig_diagram_Mn-U}~(a) presents the minority spin band structures around $\Gamma$ point calculated by 
standard GGA and the GGA+U$_\mathrm{Mn}$ with small ($U_\mathrm{eff}^\mathrm{Mn}=3$~eV) and large (6~eV) 
parameter values.
The GGA results indicate that the anti-bonding Co $t_{2g}^*$ is dominant just below the Fermi
energy and bonding Mn $t_{2g}$ is visible at $-1.5$~eV in a minority state.
Interestingly, %applying 
increasing the $U_\mathrm{eff}^\mathrm{Mn}$ value modifies the spectral weights of the minority components; 
the Mn and Co orbital weights in bonding and anti-bonding states are almost identical at small 
$U_\mathrm{eff}^\mathrm{Mn}$, but the anti-bonding $t_{2g}^*$ becomes dominant by 
Mn compared to Co and $t_{2g}^*$ shifts above the Fermi energy at a large $U_\mathrm{eff}^\mathrm{Mn}$.
Schematic diagrams are illustrated in
Figs.~\ref{fig_diagram_Mn-U}~(b) and (c). 
In the majority spin, the valence Mn $t_{2g}$ atomic orbital is shifted to a lower energy by the 
$U_\mathrm{eff}^\mathrm{Mn}$ effect and the anti-bonding state Co $t_{2g}$ is drawn to lower energy by 
the hybridization with Mn $t_{2g}$.
On the other hand, in the minority state, the energy level of the Mn valence state becomes higher 
as $U_\mathrm{eff}^\mathrm{Mn}$ increases, and the anti-bonding Co $t_{2g}^*$ orbital 
gradually touches the Fermi energy.
When the Mn $t_{2g}$ state becomes energetically higher than Co at a large $U_\mathrm{eff}^\mathrm{Mn}$,
the component of the anti-bonding $t_{2g}^*$ is switched from Co to Mn in minority spin.

%... Fig.5
\begin{figure*}
\begin{center}
\includegraphics[width=1.5\columnwidth]{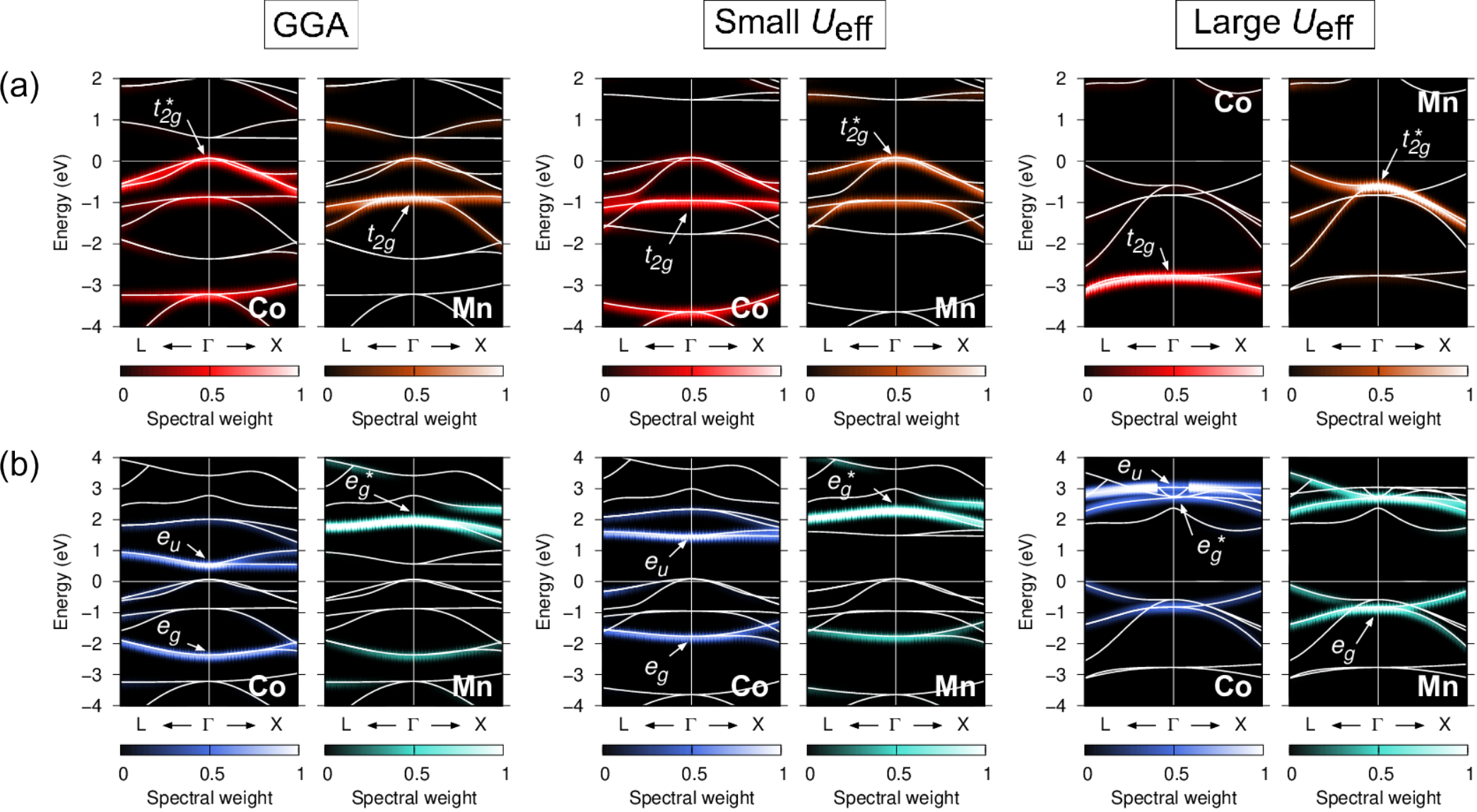}
\\~\\
\includegraphics[width=1.46\columnwidth]{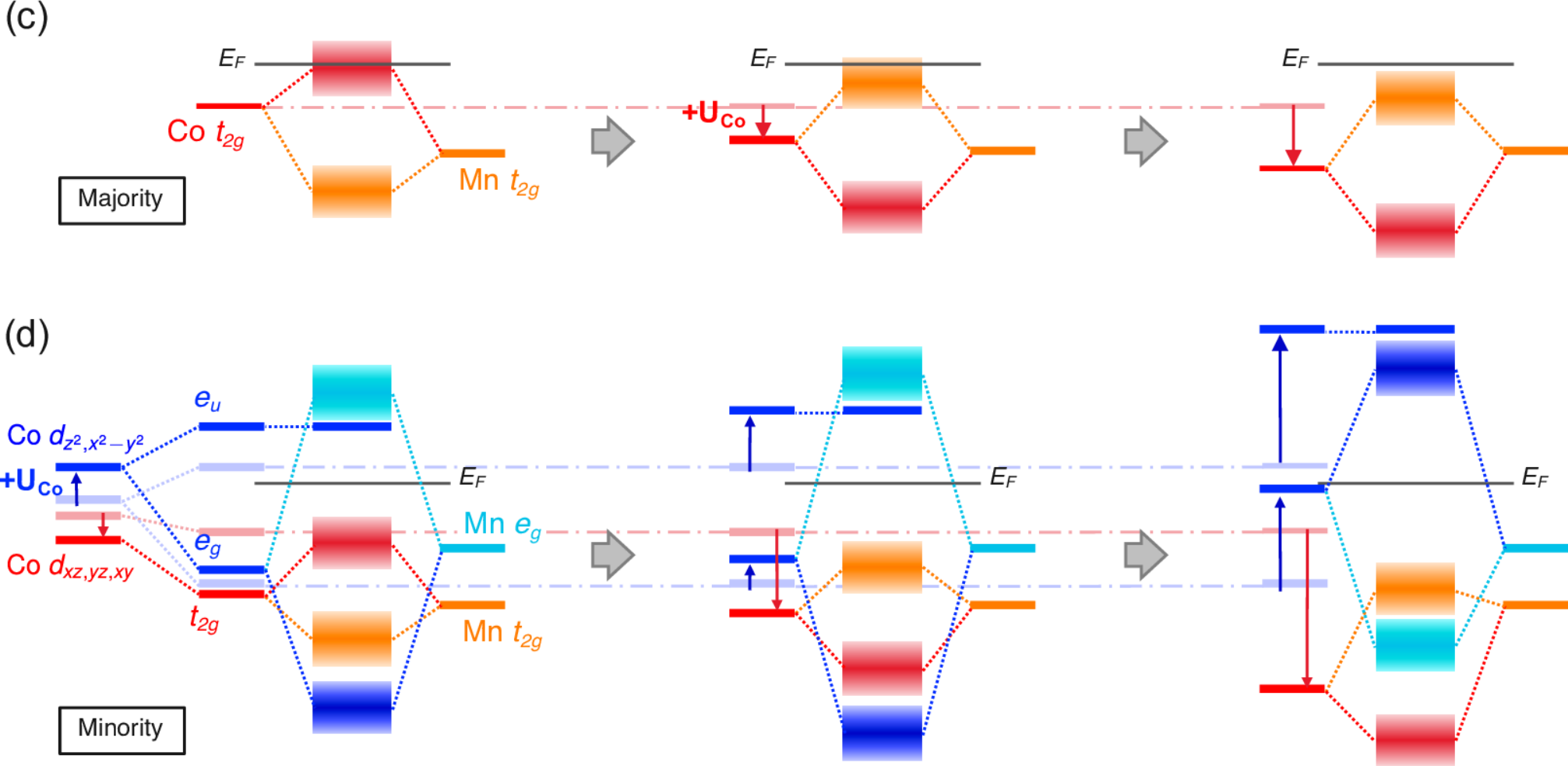}
\end{center}
\caption{(Color online) 
         Dependence of the minority band structures of varying +U$_\mathrm{Co}$ parameter, i.e.,
         GGA ($U_\mathrm{eff}=0$~eV for Co site), small (3~eV), and large (6~eV) values, where the projected 
         spectral weights for
         (a)~Co and Mn $t_{2g}$ states in the left (red) and right (orange) panels, and 
         (b)~Co and Mn $e_{g}$  ones  in the left (blue) and right (sky-blue), respectively.
         Note that the energy ranges in (a) and (b) are different.
         Schematic summary of changes in atomic orbital hybridizations for (c)~majority and (d)~minority states.
         Notation is the same in Fig.~\ref{fig_diagram_Mn-U}.}
\label{fig_diagram_Co-U}
\end{figure*}

Surprisingly, $U_\mathrm{eff}^\mathrm{Mn}$ shifts the minority occupied state of 
Mn {\it upward} energetically;
this shifting is an opposite tendency to the well-known fact of DFT+U study.
In general, the +U term opens the band gap with the valence (conduction) state being lower (higher) energy in the 
insulating and semi-conducting materials regardless of the spin channels.
However, ferromagnetic materials, including the Heusler alloy, are different from insulators and semi-conductors
because a finite DOS lies at the Fermi energy in ferromagnets.
In principle, the total number of valence electrons at each atom site must be preserved even though the +U effect
is introduced.
Accordingly, the upward shifting in the valence state of the minority Mn $d$ orbital can be understood as follows:
the occupations in a spin channel (majority state) vary to increase by the applied +U effect, but 
simultaneously, the occupations in the opposite spin (minority state) are also changed to reduce sensitively for 
the total occupations constant at each atom.
This argument is based on the energy diagrams in Figs.~\ref{fig_diagram_Mn-U}~(b) and (c), and
is consistent with the behaviors of spin magnetic moment and electron occupations at each atomic site 
[Figs.~\ref{fig_mspin-ueff}~(a) and (b)]; 
thus, this scenario can be concluded to be behind the effects of +U on Mn $d$ orbitals.

Second, the GGA+U$_\mathrm{Co}$ result is considered.
In the range of $U_\mathrm{eff}^\mathrm{Co}$ less than around 4~eV, an increase in the spin magnetic moment 
at the Co site is not significant,
but suddenly increases afterward [Fig.~\ref{fig_mspin-ueff}~(c)].
In Fig.~\ref{fig_diagram_Co-U}, the $U_\mathrm{eff}^\mathrm{Co}$-dependent electronic band structures and
hybridization behaviors of Co and Mn are summarized.
In the valence states, it can be seen that the contribution to the anti-bonding $t_{2g}^*$ is switched from Co
to Mn [Fig.~\ref{fig_diagram_Co-U}~(a)]; in contrast, the anti-bonding $e_g^*$ state is switched from Mn to Co
[Fig.~\ref{fig_diagram_Co-U}~(b)] with increasing $U_\mathrm{eff}^\mathrm{Co}$.
The $e_u$ state in Fig.~\ref{fig_diagram_Co-U}~(b) moves to higher energy by U$_\mathrm{eff}^\mathrm{Co}$, 
but does not hybridize with Mn.

To understand the behavior of changing $m_\mathrm{spin}$ and electron numbers
in GGA+U$_\mathrm{Co}$ in 
Figs.~\ref{fig_mspin-ueff}~(c) and (d), the possible energy diagrams for 
the majority and minority states are illustrated in Figs.~\ref{fig_diagram_Co-U}~(c) and (d).
The majority Co $t_{2g}$ simply goes to lower energy by the introduction of $U_\mathrm{eff}^\mathrm{Co}$, 
so that the Co $d$ spin-up
occupation increases and is saturated at larger $U_\mathrm{eff}^\mathrm{Co}$ values ($\sim7$~eV).
For minority state, the $d$  
bands' behaviors of Co and Mn are intricate, but it can be understood 
by going back to the principle view that first focus is paid to the hybridization 
between Co atoms at different sublattices and that between Mn and Co--Co states afterward,
as discussed in Sec.~\ref{subsec:org-HM} and a previous report~\cite{prb-66-174429-2002}.
The Co $d_{z^2}$ and $d_{x^2-y^2}$ ($d_{xz}, d_{yz}$, and $d_{xy}$) orbitals are pushed up (down) due to 
$U_\mathrm{eff}^\mathrm{Co}$, and hybridize with Mn $e_g$ ($t_{2g}$) state [Fig.~\ref{fig_diagram_Co-U}~(d)].
Increasing $U_\mathrm{eff}^\mathrm{Co}$ affects the energy gap, and most notably, the Co $e_g$ orbital 
becomes an un-occupied anti-bonding state at a large $U_\mathrm{eff}^\mathrm{Co}$ value, 
while it is an occupied bonding state at a small 
$U_\mathrm{eff}^\mathrm{Co}$ (see blue band of energy diagram in Fig.~\ref{fig_diagram_Co-U}~(d)).
This event induces a significant reduction in the minority Co occupations 
[red down-pointing triangle in Fig.~\ref{fig_mspin-ueff}~(d)],
resulting in an increase in the total $m_\mathrm{spin}$ in the range over %$\sim6$~eV 
$\sim4$~eV of $U_\mathrm{eff}^\mathrm{Co}$, as shown by black plots in Fig.~\ref{fig_mspin-ueff}~(c).

From the above discussions,
the underlying physics of the correlation effects on the magnetic moment can be addressed 
from the viewpoint of electronic structure for both GGA+U$_\mathrm{Mn}$ and GGA+U$_\mathrm{Co}$ cases.
Thus, the consistency of our energy diagram proposed in Fig.~\ref{fig_band_gga}~(c) is demonstrated
successfully.

%-- Electronic and magnetic properties ------------------------------------------------------------------
\subsection{Electronic and magnetic properties}\label{subsec:el-mag}

We now discuss the electronic and magnetic properties obtained from the band calculations that incorporate the 
LR-determined correlation parameters (3.535~eV for Mn and 6.570~eV for Co).
First, the $m_\mathrm{spin}$ obtained from the GGA+U$_\mathrm{Mn}^\mathrm{LR}$ method is compared with the 
GGA in Table~\ref{table1}.
The value of total $m_\mathrm{spin}$ is same as that of GGA and agrees with previous reports.~\cite{jmmm-422-13-2017,
prb-88-134402-2013,jphysDapplphys-42-084002-2009,prb-86-245115-2012,
prb-74-104405-2006,prb-80-144405-2009,jpcs-32-1221-1971,jmmm-937-54-1986}
On the other hand, 
regarding the atom-resolved contributions, the results of $m_\mathrm{spin}$ of Co (0.72~$\mu_\mathrm{B}$) 
and Mn (3.63~$\mu_\mathrm{B}$) are not in agreement with GGA and previous calculations, 
but in good agreement with the experiments~\cite{jpcs-32-1221-1971,prb-80-144405-2009}.
Thus, the GGA+U$_\mathrm{Mn}^\mathrm{LR}$ calculation results are superior to the standard GGA results.

%... Fig.8
\begin{figure}
\begin{center}
\includegraphics[width=0.9\columnwidth]{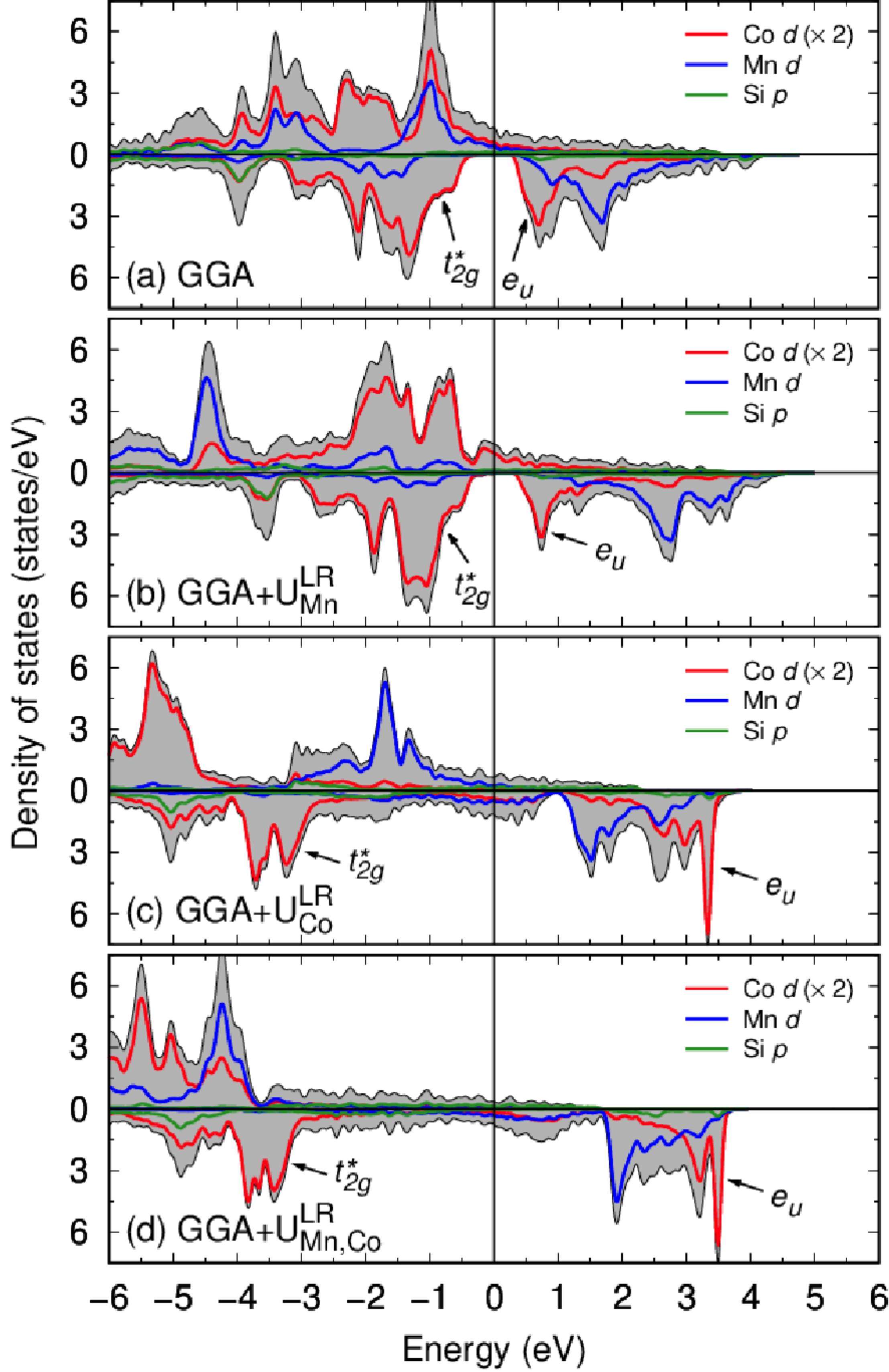}
\end{center}
\caption{(Color online) 
         Local DOS obtained from (a)~GGA, (b)~GGA+U$_\mathrm{Mn}^\mathrm{LR}$, 
         (c)~GGA+U$_\mathrm{Co}^\mathrm{LR}$, and (d)~GGA+U$_\mathrm{Mn,Co}^\mathrm{LR}$ calculations.
         Red, blue, and green lines are for Co $d$, Mn $d$, and Si $p$ orbitals, and
         total DOS is shown by a gray filled area.
         The orbital characters of $t_{2g}^*$ and $e_u$ states, which originate from Co, are shown with arrows. 
         Note that the local DOS for Co is twice as two Co atoms are included in the primitive cell.
         The upper (bottom) area in each panel shows the spin-up (-down) state, and the Fermi energy is 
         set to zero.}
\label{fig_dos_CMS}
\end{figure}

In the GGA-calculated DOS in Fig.~\ref{fig_dos_CMS}~(a),
we can clearly see that the Co $d$ orbital is broad over a wide energy region 
(from Fermi energy to $-5$~eV for majority state and from $-0.5$~eV to $-4.5$~eV for minority state).
Contrarily, the Mn $d$ orbital is relatively localized compared to the Co one and splits into two peaks 
located around $-3$ and $-1$~eV ($-1.5$ and $1.8$~eV) in majority (minority) state, respectively.
As expected from Figs.~\ref{fig_band_gga}~(b) and (c), 
we also confirm the $e_u$ and $t_{2g}^*$ orbital characters of Co $d$ states above and below the Fermi energy,
as shown by arrows in Fig.~\ref{fig_dos_CMS}~(a).
The value of spin polarization referred as $P_\mathrm{DOS}$ is estimated by
$P_\mathrm{DOS}=\frac{D^\uparrow(E_\mathrm{F})-D^\downarrow(E_\mathrm{F})}
                     {D^\uparrow(E_\mathrm{F})+D^\downarrow(E_\mathrm{F})} \times 100 ~(\%)$,
where $D^\sigma(E_\mathrm{F})$ is the DOS of the majority ($\sigma=\uparrow$) or minority ($\sigma=\downarrow$) 
spin state at the Fermi energy.
A 100~\% $P_\mathrm{DOS}$ value is obtained, and the energy band gap in the minority spin state 
$E_\mathrm{gap}^\downarrow$ is around 0.8~eV.
The GGA+U$_\mathrm{Mn}^\mathrm{LR}$ calculation modifies the DOS from GGA.
The energy level of Mn occupied (unoccupied) states is shifted to lower (higher) level due to the exchange 
splitting induced by the $U_\mathrm{eff}^\mathrm{LR(Mn)}$ effect.
As a result, the valence and conduction edges are dominated mainly by Co $d$ components
and only a few Mn $d$ states appear around the Fermi energy.
Due to the presence of a few Co $d$ DOSs at the Fermi energy, 
the half-metallicity is broken but high spin polarization $P_\mathrm{DOS}=90.5~\%$ is obtained.

By contrast, GGA+U$_\mathrm{Co}^\mathrm{LR}$ and GGA+U$_\mathrm{Mn,Co}^\mathrm{LR}$ seem to fail 
to obtain the total magnetic moment reasonably consistent with the experimental
observations~\cite{jpcs-32-1221-1971,prb-80-144405-2009} because of the overestimated 
$U_\mathrm{eff}^\mathrm{LR(Co)}$ parameters (see Table~\ref{table1}).
Figure~\ref{fig_dos_CMS}~(c) indicates the fact that the exchange splitting arising from the large 
$U_\mathrm{eff}^\mathrm{LR(Co)}$ induces a fully-occupied Co $d$ state in majority spin states, 
which leads to a Co $m_\mathrm{spin}$ of 1.88~$\mu_\mathrm{B}$ and total $m_\mathrm{spin}$ of 6.95~$\mu_\mathrm{B}$.
The energy gap does not appear in the minority channel and the top of valence states around $-2$~eV from 
Fermi energy is composed of the Mn $d$ orbital of majority states.
Similarly, in the GGA+U$_\mathrm{Mn,Co}^\mathrm{LR}$ case, the overestimated value of the total $m_\mathrm{spin}$ of 
$8.08~\mu_\mathrm{B}$ arises from that the majority electrons of Co and Mn are fully occupied
at low energy ($-4$~eV and below) 
through both $U_\mathrm{eff}^\mathrm{LR(Mn)}$ and $U_\mathrm{eff}^\mathrm{LR(Co)}$,
as shown in Fig.~\ref{fig_dos_CMS}~(d).
In this scheme,
the half-metallic electronic structure is broken by a few DOS that is widely broad around the Fermi energy.
The spin polarizations are found to be negative and small absolute values, $P_\mathrm{DOS}=-26.97$ and $-33.82$~\%
for the respective GGA+U$_\mathrm{Co}^\mathrm{LR}$ and GGA+U$_\mathrm{Mn,Co}^\mathrm{LR}$ methods.

From the experimental viewpoint,
the hard X-ray photoelectron spectroscopy measurements reported that
the valence band structure in the binding energy region from Fermi energy to $\sim1.2$~eV
(corresponding to $-1.2$~eV in calculated DOS) is 
mostly contributed by Co $3d$ electrons~\cite{prb-79-100405R-2009} 
and the Mn $d$ state does exist in this binding-energy region, while the
number of electrons is very few compared to Co.~\cite{prv1}
Based on the above comparative discussions between our calculations and experiments on the electronic structure and 
magnetic moment (as well as the equilibrium lattice constant in Sec.~\ref{subsec:str}),
we can conclude that the static many-body correlation +U at $Y$ site ($Y=\rm{Mn}$ for Co$_2$MnSi)
plays an important role in ground-state properties that are in good agreement with the experiments.
On the other hand, the Co $d$ electrons are rather itinerant in the alloy; thus, the LR approach tends to 
overestimate the correlation parameter for Co site, which is not reliable for accurate 
band calculations.
In other words, for Co site, correlation correction may not be necessary and mean-field approximation 
(GGA or LSDA) is enough to treat the itinerant Co $d$ electrons.
Thus, hereafter, all LR-based DFT+U calculations are performed with $U_\mathrm{eff}^\mathrm{LR}$ only for $Y$ site;
i.e., correlation correction is excluded for Co.
We here explicitly mention that the energy diagram obtained from the GGA+U$_\mathrm{Mn}^\mathrm{LR}$ calculation
corresponds to Figs.~\ref{fig_band_gga}~(b) and (c), which are obtained from the GGA results.

%-- Search for HM materials of other ternary and quaternary alloys --------------------------------------
\section{Search for HM materials of other ternary and quaternary alloys}\label{sec:CMYS}

%... Fig.9
\begin{figure*}[t]
\begin{center}
\includegraphics[width=1.8\columnwidth]{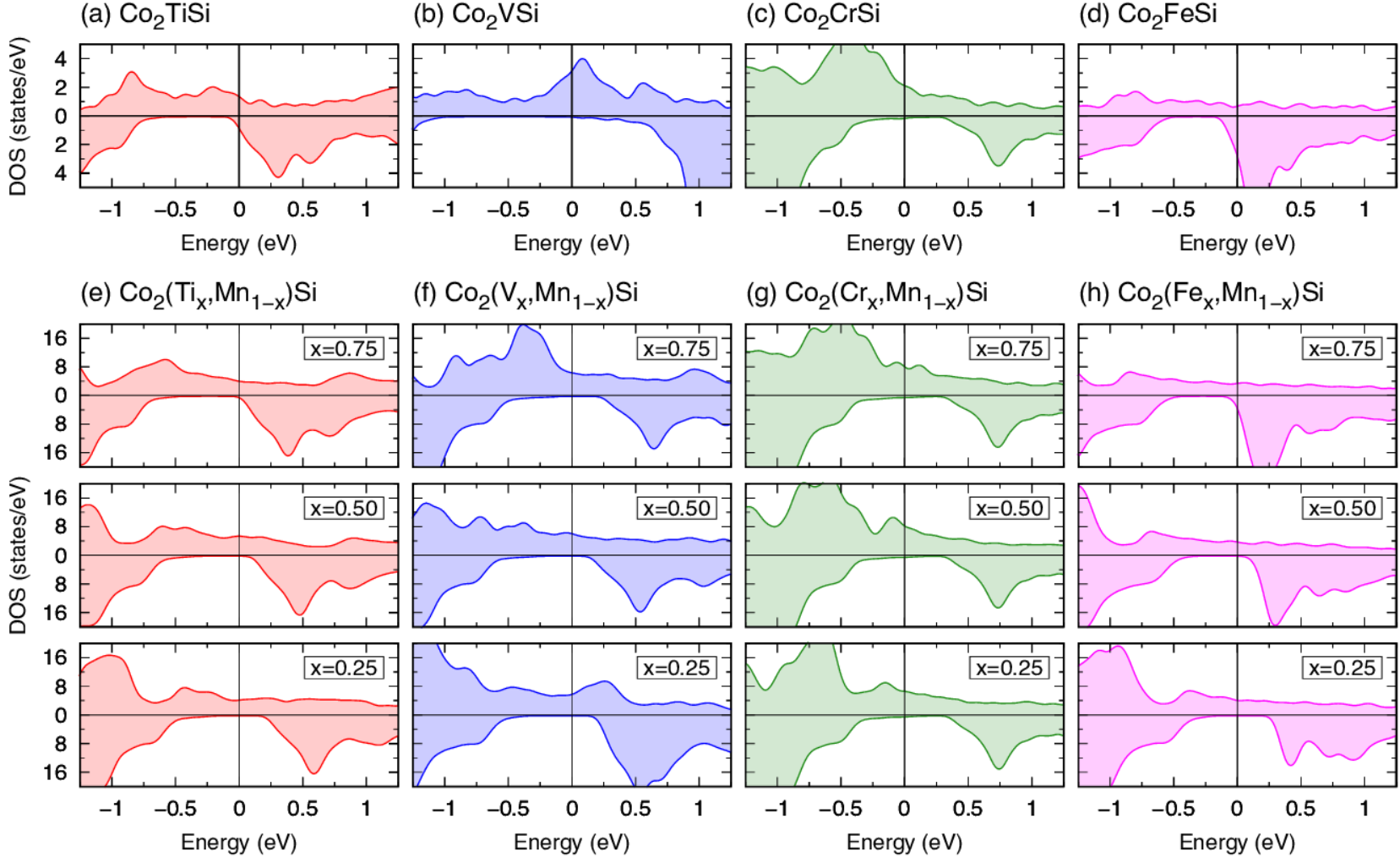}
\end{center}
\caption{(Color online) 
         (a)$\sim$(d)~Total DOS for ternary Co$_2Y$Si ($Y=$ Ti, V, Cr, or Fe), and
         (e)$\sim$(h)~total DOS dependence on composition $x$ for quaternary Co$_2$($Y_x$,Mn$_{1-x}$)Si
         ($x=0.25$, 0.5, or 0.75) calculated by the LR-based DFT+U method.
         In each panel, the upper (bottom) region shows the DOS for up- (down-) spin state,
         and the Fermi energy is set to zero.
         Note that the vertical axis range of DOS in (a)$\sim$(d) is different from that in (e)$\sim$(h)  
         as the number of atoms per primitive cell of the ternary system is a quarter of the quaternary one
         (see Fig.~\ref{fig_model}).}
\label{fig_dos_CMYS}
\end{figure*}

For considering the ternary Co$_2Y$Si alloys, where $Y$ is replaced from Mn to Ti, V, Cr, or Fe atom, 
the LR calculations for $U_\mathrm{eff}^{\mathrm{LR(}Y\mathrm{)}}$ are first carried out 
using the lattice constants assumed in the experiments~\cite{jmmm-38-1-1983,apl-88-032503-2006,spmmc1-262-1962},
as in the case of Co$_2$MnSi.
To the best of our knowledge, there are no experimental data for Co$_2$CrSi, so the lattice constant obtained
from Murnaghan fitting~\cite{pnasusa-30-244-1944} by the GGA potential is employed for 
$U_\mathrm{eff}^\mathrm{LR(Cr)}$ calculation.
In this study initial magnetization for the SCF calculation is assumed to the ferromagnetic state in 
all ternary models.
The determined parameters are around $3\sim4$~eV depending on the materials;
the $U_\mathrm{eff}^{\mathrm{LR(}Y\mathrm{)}}=2.942,$ 3.979, 3.169, and 3.922~eV for 
Co$_2$TiSi, Co$_2$VSi, Co$_2$CrSi, and Co$_2$FeSi, respectively.~\cite{note_ulrt_Co2YSi}

Calculated total DOSs are shown in Figs.~\ref{fig_dos_CMYS}~(a)$\sim$(d),
and the results of spin magnetic moments are summarized in Table~\ref{table2}.
The Co$_2$TiSi is not the HM ($P_\mathrm{DOS}=25.8~\%$), where the Fermi energy is located at the minority conduction edge state.
For the Co$_2$VSi and Co$_2$CrSi, a few broad minority DOSs are found around the Fermi energy;
thus, the electronic structure is not HM, but the highly spin-polarized values are estimated as $P_\mathrm{DOS}=98.2$
and 89.3~\%, respectively.
On the other hand, negative spin polarization, $P_\mathrm{DOS}=-62.3~\%$, is obtained in the Co$_2$FeSi, 
where the minority DOS is much compared to the majority state at Fermi energy.
Note that, as the $Y$ atom is changed from a large atomic number ($Z_\mathrm{Fe}=26$) to small 
($Z_\mathrm{V}=23$), the Fermi energy position seems to move away from the conduction state of minority spin,
but this is not the case for $Y=\mathrm{Ti}$.
This exception is attributable to the fact that the Ti spins in Co$_2$TiSi couple with those of Co with anti-parallel
direction and the ferrimagnetic structure is obtained in our calculations, while the other systems favor the
ferromagnetic structure (see Table~\ref{table2}).
Total spin magnetic moments are calculated as 1.89, 3.00, 4.03, and 5.42~$\mu_\mathrm{B}$ for
Co$_2$TiSi, Co$_2$VSi, Co$_2$CrSi, and Co$_2$FeSi, respectively.

%... TABLE 3
\begin{table} 
\caption{Nominal number of valence electrons $N_\mathrm{val}$ and calculated spin magnetic moments of 
         total and atom-resolved contributions (in unit of $\mu_\mathrm{B}$)
         for  Co$_2Y$Si ($Y=$ Ti, V, Cr, or Fe) and 
          Co$_2$($Y_x$,Mn$_{1-x}$)Si ($x=$ 0.25, 0.50, or 0.75).
         Results are obtained from the LR-based DFT+U method.
         }
\begin{threeparttable}
\begin{ruledtabular}
\renewcommand{\arraystretch}{1.3}
\begin{tabular}{llccrc}
 & \multirow{2}{*}{$N_\mathrm{val}$}                         &
   \multicolumn{4}{c}{Spin magnetic moment} \\
\cline{3-6}
 & & Total & Co & $Y~$ & Mn \\
\hline
%                                   Nval    total    Co      Y         Mn    
Co$_2$TiSi                       &  26     & 1.89  & 0.97 & $-0.02$  &       \\
Co$_2$(Ti$_{0.75}$Mn$_{0.25}$)Si &  26.75  & 2.76  & 0.94 & $-0.09$  & 3.85  \\
Co$_2$(Ti$_{0.50}$Mn$_{0.50}$)Si &  27.5   & 3.54  & 0.90 & $-0.21$  & 3.75  \\
Co$_2$(Ti$_{0.25}$Mn$_{0.75}$)Si &  28.25  & 4.28  & 0.83 & $-0.37$  & 3.69  \\
\hline
Co$_2$VSi                        &  27     & 3.00  & 1.26 & $ 0.59$  &       \\
Co$_2$(V$_{0.75}$Mn$_{0.25}$)Si  &  27.5   & 3.51  & 0.84 & $ 1.19$  & 3.75  \\
Co$_2$(V$_{0.50}$Mn$_{0.50}$)Si  &  28     & 4.03  & 0.38 & $ 1.06$  & 3.70  \\
Co$_2$(V$_{0.25}$Mn$_{0.75}$)Si  &  28.5   & 4.56  & 1.10 & $ 0.15$  & 3.13  \\
\hline
Co$_2$CrSi                       &  28     & 4.03  & 0.52 & $ 2.90$  &       \\
Co$_2$(Cr$_{0.75}$Mn$_{0.25}$)Si &  28.25  & 4.28  & 0.58 & $ 2.90$  & 3.63  \\
Co$_2$(Cr$_{0.50}$Mn$_{0.50}$)Si &  28.5   & 4.52  & 0.62 & $ 2.88$  & 3.61  \\
Co$_2$(Cr$_{0.25}$Mn$_{0.75}$)Si &  28.75  & 4.77  & 0.67 & $ 2.94$  & 3.63  \\
\hline
Co$_2$(Fe$_{0.25}$Mn$_{0.75}$)Si &  29.25  & 5.27  & 0.94 & $ 2.91$  & 3.65  \\
Co$_2$(Fe$_{0.50}$Mn$_{0.50}$)Si &  29.5   & 5.55  & 1.15 & $ 2.94$  & 3.69  \\
Co$_2$(Fe$_{0.75}$Mn$_{0.25}$)Si &  29.75  & 5.58  & 1.25 & $ 2.95$  & 3.75  \\
Co$_2$FeSi                       &  30     & 5.42  & 1.29 & $ 2.92$  &       \\
\end{tabular}
\end{ruledtabular}
\end{threeparttable}
\label{table2}
\end{table}

The structural properties are also investigated as summarized in Table~\ref{table:str_CMYS}.
The estimated lattice constants are in good agreement with 
the experiments~\cite{jmmm-38-1-1983,apl-88-032503-2006,spmmc1-262-1962,jap-100-083523-2011}
and their error values from the experiments are less than 1~\% for Co$_2$TiSi, Co$_2$VSi, and Co$_2$FeSi.
In Co$_2$CrSi, the lattice constant of 5.694~{\AA} is close to the previous 
calculation.~\cite{jap-100-113901-2006}
The bulk moduli in all models estimated from the LR-based DFT+U method are slightly smaller than those in the previous 
calculations.
This trend is similar to the Co$_2$MnSi case, and might come from that the previous studies were conducted by
standard LSDA\cite{physicaB-407-3339-2012,jap-100-113901-2006,jap-125-082523-2019,UnvJMechanEng-7-16-2019,
jpcs-75-391-2014} and GGA~\cite{IEEETransMagn-50-1301104-2014}.
The experimentally measured $B_0$ is available only for $Y=$ Fe ($B_0=240$~GPa).~\cite{jap-100-083523-2011}
From our calculations, the $B_0$ and $B_0'$ in Co$_2$FeSi are found to be 183.263~GPa and 4.679, respectively.
The LSDA calculation~\cite{IEEETransMagn-50-1301104-2014} shows a reasonably consistent value of 
$B_0=241.9$~GPa with the experiment, although the GGA 
calculation~\cite{jap-100-113901-2006,jap-125-082523-2019,jpcs-75-391-2014} 
underestimates $B_0$ ($B_0=203.5\sim 207.1$~GPa).
Zhu {\it et al.}~\cite{jpcs-75-391-2014} also performed the GGA+U calculations, where the empirical 
parameters of $U=3.5$ and $J=0.9$~eV for Co and those of $U=3.4$ and $J=0.9$~eV for Fe are employed, 
and obtained $B_0=209.3$~GPa and $B_0'=4.67$ (the GGA+U results are not shown in Table~\ref{table:str_CMYS}).
Therefore, the LSDA calculations might be suitable for the bulk modulus compared to the GGA+U approaches,
while it seems to underestimate the lattice constant from the experiments, 
for example, $a_0=5.52$~{\AA} in Co$_2$FeSi~\cite{IEEETransMagn-50-1301104-2014}.
However, the LR-based DFT+U method provides reasonable results at least for $a_0$ values.

%... TABLE 4
\begin{table*} 
\caption{Structural parameters of lattice constant $a_0$, bulk modulus $B_0$, and its pressure derivative $B_0'$ 
         for Co$_2Y$Si compared with the present study, previous calculations, and experiments.
         Results of the present study are obtained from the LR-based DFT+U method with the parameters of 
         $U_\mathrm{eff}^{\mathrm{LR(}Y\mathrm{)}}=2.942,$ 3.979, 3.169, and 3.922~eV for $Y=$ Ti, V, Cr, and Fe,
         respectively.
         Previous calculation results are from GGA, except for the bottom row for Co$_2$FeSi that are from LSDA.}
\begin{threeparttable}
\begin{ruledtabular}
\renewcommand{\arraystretch}{1.3}
\begin{tabular}{lcccccccc}
        & \multicolumn{3}{c}{Present work}
        & \multicolumn{3}{c}{Theory}
        & \multicolumn{2}{c}{Experiment}       \\
\cline{2-4} \cline{5-7} \cline{8-9}
        & $a_0$~(A) & $B_0$~(GPa) & $B_0'$        
        & $a_0$~(A) & $B_0$~(GPa) & $B_0'$
        & $a_\mathrm{Expt}$~(A) & $B_0$~(GPa)   \\
\hline
Co$_2$TiSi&5.774&189.494&4.191&5.764\tnote{a}                   & 204--244.8304\tnote{b,a,e,c}  & 4.5151\tnote{e} & 5.743\tnote{g}     &  \\
Co$_2$VSi &5.667&192.408&7.485&5.7609\tnote{b} , 5.679\tnote{c} & 216\tnote{b} , 221.5\tnote{c} &                 & 5.647\tnote{h}     &  \\
Co$_2$CrSi&5.694&174.169&5.106&5.6295\tnote{b} , 5.638\tnote{c} & 227\tnote{b} , 225.3\tnote{c} &                 &                    &  \\
Co$_2$FeSi&5.685&183.263&4.679&5.6431\tnote{b}                  & 203.5--207.1\tnote{b,f,c}     & 4.62\tnote{f}   & 5.644\tnote{i}, 5.650\tnote{j} & 240\tnote{j}\\
          &     &       &     &5.52\tnote{d}                    & 241.9\tnote{d}                &                 &                    & \\  
\end{tabular}
\begin{tablenotes}[para]
  \item[a] Reference~[\onlinecite{physicaB-407-3339-2012}]
  \item[b] Reference~[\onlinecite{jap-100-113901-2006}]
  \item[c] Reference~[\onlinecite{jap-125-082523-2019}]
  \item[d] Reference~[\onlinecite{IEEETransMagn-50-1301104-2014}]
  \item[e] Reference~[\onlinecite{UnvJMechanEng-7-16-2019}]
  \item[f] Reference~[\onlinecite{jpcs-75-391-2014}]
  \item[g] Reference~[\onlinecite{jmmm-38-1-1983}]
  \item[h] Reference~[\onlinecite{apl-88-032503-2006}]
  \item[i] Reference~[\onlinecite{spmmc1-262-1962}]
  \item[j] Reference~[\onlinecite{jap-100-083523-2011}]
\end{tablenotes}
\end{ruledtabular}
\end{threeparttable}
\label{table:str_CMYS}
\end{table*}

We finally investigate the quaternary Heusler compounds of a chemical formula Co$_2$($Y_x$,Mn$_{1-x}$)Si
($Y=\mathrm{Ti}$, V, Cr, or Fe) with a composition $x$ ($x=0.25, 0.50$, or 0.75).
To model these systems in $L2_1$ structure, as illustrated in Figs.~\ref{fig_model}~(b)$\sim$(d), cubic primitive 
cells consisting of 16 atoms are considered.
The lattice constant is given by Vegard's 
law~\cite{ZPhysik-5-17-1920,pra-43-3161-1991} using the obtained equilibrium lattice constants for 
Co$_2Y$Si ($a^{\mathrm{C}Y\mathrm{S}}$) and Co$_2$MnSi ($a^\mathrm{CMS}$) as 
$a(x) = x a^{\mathrm{C}Y\mathrm{S}} + (1-x) a^\mathrm{CMS}$.
The correlation parameters of Mn and $Y$ atoms for quaternary systems at all compositions are assumed to 
be the values of 
$U_\mathrm{eff}^\mathrm{LR(Mn)}$ and $U_\mathrm{eff}^{\mathrm{LR(}Y\mathrm{)}}$, which are determined by the LR 
theory in ternary Co$_2$MnSi and Co$_2Y$Si.
For the quaternary compounds, 
in which the atomic position of the different elements is not symmetric as the ternary system,
the structures are geometrically relaxed under the equilibrium lattice constants by force calculations 
using the Broyden-Fletcher-Goldfarb-Shanno (BFGS) algorithm~\cite{SIAP-6-76-291970,ComputJ-J13-317-1970,
MathComputModell-24-23-1970,MathComputModell-24-647-1970}
until the forces acting on each atom are minimized below the criterion of $10^{-3}$~Ry/bohr.

\begin{figure}
\begin{center}
\includegraphics[width=0.95\columnwidth]{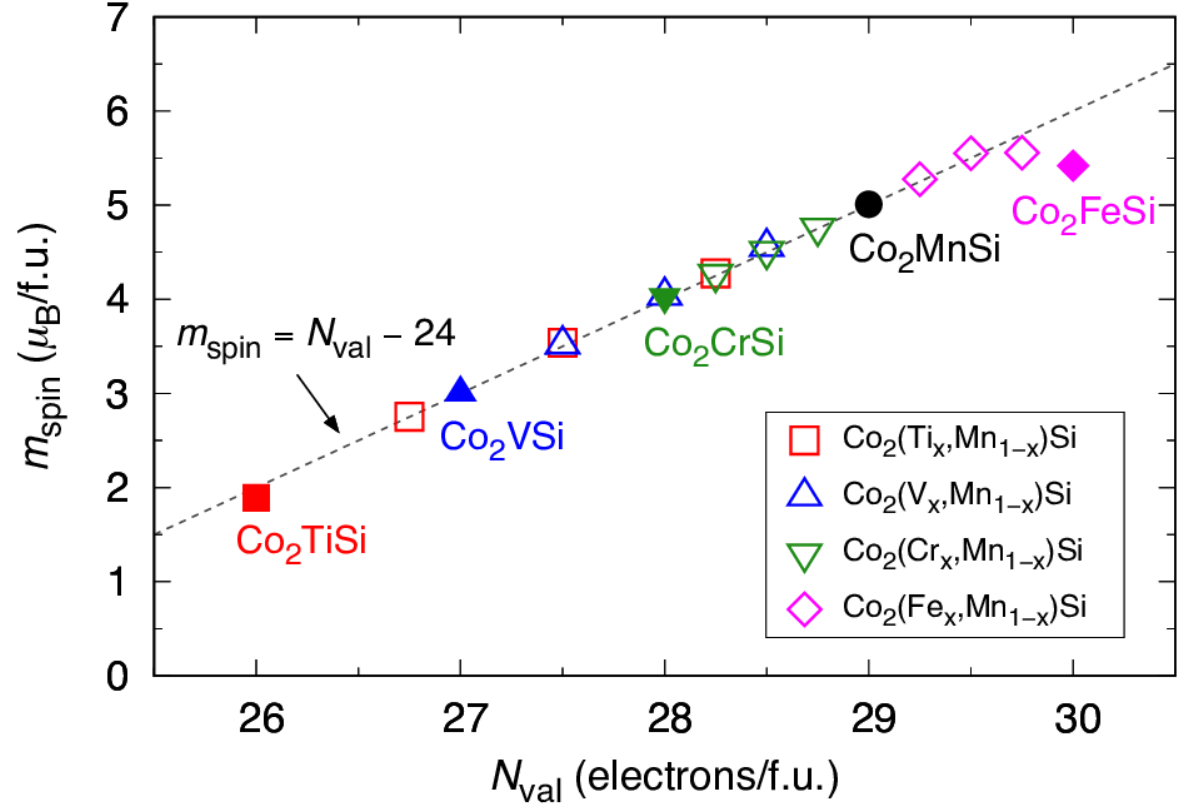}
\end{center}
\caption{(Color online) 
         Total $m_\mathrm{spin}$ as a function of $N_\mathrm{val}$ for 
         ternary Co$_2Y$Si ($Y=$ Ti, V, Cr, or Fe)
         and quaternary Co$_2$($Y_x$,Mn$_{1-x}$)Si ($x=$ 0.25, 0.5, or 0.75).
         Closed square (red), triangle (blue), down-pointing triangle (green), and diamond (pink)
         indicate ternary systems, respectively.
         Opened symbols are for quaternary systems.
         The results of Co$_2$MnSi are also plotted by a closed circle (black).
         The Slater-Pauling relation, $m_\mathrm{spin}=N_\mathrm{val}-24$, is shown by a dotted line.}
\label{fig_mspin_nval}
\end{figure}

The calculated $m_\mathrm{spin}$ values for quaternary alloys are also available in Table~\ref{table2}.
In case of only $Y=\mathrm{Ti}$, the Ti spins are ferrimagnetically coupled with Co and Mn similar to that 
in the ternary model.
The Mn $m_\mathrm{spin}$ is a large value over 3~$\mu_\mathrm{B}$ in all systems. 
Figure~\ref{fig_mspin_nval} plots the total $m_\mathrm{spin}$ for ternary and quaternary Co-based full
Heusler compounds under study as a function of $N_\mathrm{val}$ in the system. 
The Slater-Pauling relation is satisfied in the range of less than 29.5 in $N_\mathrm{val}$ while being slightly
underestimated for the range over $N_\mathrm{val}=29.5$, which corresponds to Co$_2$(Fe$_{0.75}$,Mn$_{0.25}$)Si, 
and Co$_2$FeSi.

%... Fig.10

\begin{figure} [b]
\begin{center}
\includegraphics[width=0.93\columnwidth]{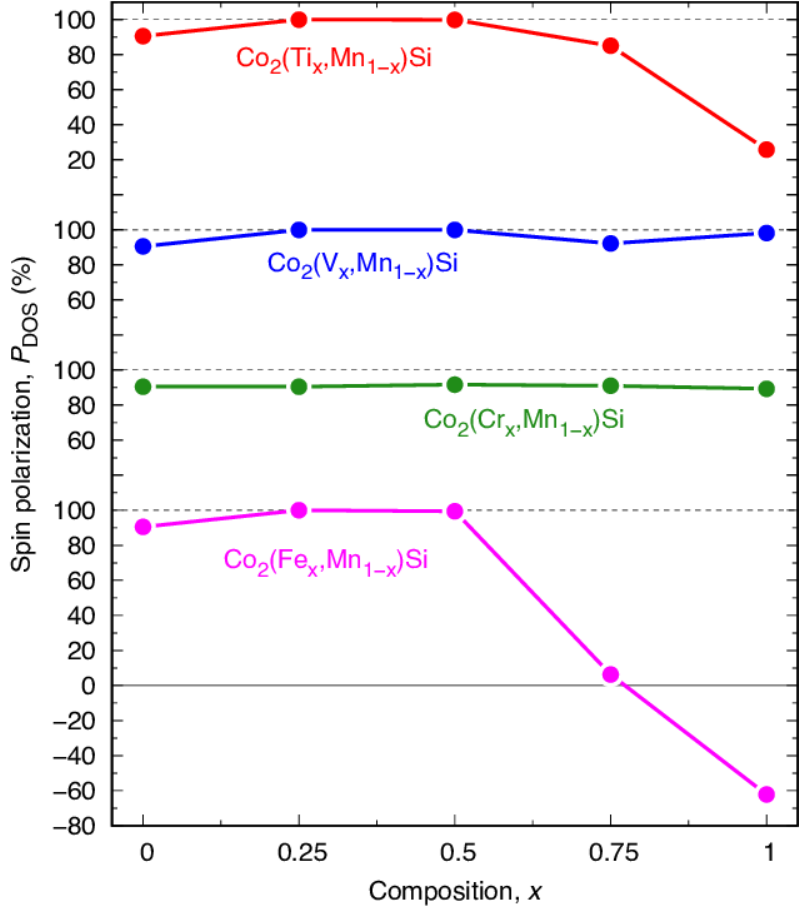}
\end{center}
\caption{(Color online) 
         Spin polarization $P_\mathrm{DOS}$ dependence on composition $x$ in Co$_2$($Y_x$,Mn$_{1-x}$)Si.
         Red, blue, green, and pink plots are $Y$ of Ti, V, Cr, and Fe, respectively.}
\label{fig_pol}
\end{figure}

The results on DOS for quaternary alloys are shown in Figs.~\ref{fig_dos_CMYS}~(e)$\sim$(h).
A perfectly HM electronic structure ($P_\mathrm{DOS}$ is equal to 100~\%) is found in 
Co$_2$(Ti$_{0.25}$,Mn$_{0.75}$)Si, Co$_2$(V$_{0.25}$,Mn$_{0.75}$)Si,
Co$_2$(V$_{0.50}$,Mn$_{0.50}$)Si, and Co$_2$(Fe$_{0.25}$,Mn$_{0.75}$)Si.
Among them, the Co$_2$(V$_{0.75}$,Mn$_{0.25}$)Si has the largest 
minority band gap $E_\mathrm{gap}^\downarrow=0.5$~eV, and
thus, this material can be a good candidate for a wide-gap HM ferromagnet.
The Co$_2$(Ti$_{0.25}$,Mn$_{0.75}$)Si and Co$_2$(Fe$_{0.25}$,Mn$_{0.75}$)Si are also HM candidates because of
the advantage in Fermi energy position that locates at almost the center of the valence and conduction states in 
minority states.
These HM characters lead to the robustness of spin polarization due to the broadening of valence and conduction
states at finite temperature.
Nearly HM ($P_\mathrm{DOS}$ is almost 100~\%) is found in 
Co$_2$(Ti$_{0.50}$,Mn$_{0.50}$)Si ($P_\mathrm{DOS}=99.9$~\%) and 
Co$_2$(Fe$_{0.50}$,Mn$_{0.50}$)Si (99.4~\%).
Figure~\ref{fig_pol} presents the composition dependence of $P_\mathrm{DOS}$.
Although the $Y=$ Cr system does not show the HM property at each composition, 
an interesting trend we observed is that a high $P_\mathrm{DOS}$ is independent of the 
composition, where $Y=$ V is also the same, whereas a large reduction of $P_\mathrm{DOS}$ occurs with an 
increase in $x$, especially in the Co(Fe$_x$,Mn$_{1-x}$)Si.

%... Fig.11

Finally, we states the results of systems including Fe by comparing the previous studies.
The Co$_2$FeSi compound is still under debate to judge whether its electronic 
structure shows HM for past few decades from theories with and without the correlation 
effects.~\cite{prb-72-184434-2005,prb-73-094422-2006}
Our LR-based DFT+U calculations indicate that it is not a HM ferromagnet.
However, we emphasize that tuning the composition in quaternary Co$_2$(Fe,Mn)Si demonstrates that the 
electronic structure can be HM.
This conclusion is supported by a consistency in anisotropic magnetoresistance (AMR)
measurement.~\cite{apl-104-172407-2014}
According to an extended model for AMR formulated by Kokado {\it et al.}, 
the negative sign of the AMR effect, arising from the empty DOS either spin-up or -down states at Fermi level,
is a signature of HM.~\cite{jpsj-81-024705-2012,amr-750-978-2013,pssc11-1026-2014}
Based on this model analysis, a positive behavior of AMR is found in the Co$_2$FeSi~\cite{apl-104-172407-2014}
that indicates a ferromagnetic without the minority band gap but the negative sign is confirmed in 
the Co$_2$(Fe,Mn)Si~\cite{prb-86-020409R-2012}, leading to HM.
Note that the composition range of Fe and Mn for Co$_2$(Fe,Mn)Si, showing HM, is different 
between our study and the AMR experiment, which may be because the present quaternary models 
[Figs.~\ref{fig_model}~(b)$\sim$(d)] are assumed to be a periodic structure missing 
a perfectly disordered property of Fe and Mn, and/or the ordering parameter of $L2_1$ structure in the
experiment~\cite{apl-104-172407-2014} is rather low at all compositions.
Nonetheless, we suggest that the quaternary Co$_2$(Fe$_{0.25}$,Mn$_{0.75}$)Si is one of the most promising 
candidates as an HM Heusler ferromagnet because of the 
sizable $E_\mathrm{gap}^\downarrow(=0.4~\mathrm{eV})$ and Fermi energy position being at almost the center of the gap.
We believe our present results encourage the experiments to improve the degree of crystallinity of bulk Heusler 
alloys and/or
to fabricate a clean interface without any atomic inter-diffusion in MTJ and CPP-GMR devices for the 
enhancement of MR performances in the future.

\section{Summary}\label{sec:summary}

In summary, we revisited the fundamental electronic structure and effects of correlation parameters for $3d$ 
electrons in a Co-based full Heusler Co$_2Y$Si alloy via the LR-based DFT+U method, where the 
correlation correction $U_\mathrm{eff}$ parameters were determined from the LR approach
and the +U formalism was incorporated as the FLL form.
Focusing on Co$_2$MnSi ($Y=$ Mn), 
we considered the origin of the minority HM gap from the projected band structures calculated by the standard GGA,
and found that the $t_{2g}$ hybridization between Co and Mn is important for the gap.
The energy diagram of atomic-orbital hybridizations revealed that the HM gap originates from Co $e_u$
of the conduction state and Co--Mn hybridizing $t_{2g}$ orbitals of the valence state at the Fermi energy.
Thus, the gap is tunable by selecting a $Y$ element and/or mixing
different elements into $Y$ site through $t_{2g}$ atomic orbital coupling.
The LR calculations tend to obtain a reasonable value as a correlation parameter for $Y$ site ($Y=$ Mn
in Co$_2$MnSi) but an unexpectedly large value for Co site, which misleads to an unphysical ground state.
The failure in determining $U_\mathrm{eff}^\mathrm{LR}$ for Co site arises from the fact that the $d$ electrons of 
Co site behave rather itinerant in the alloy.
This means that the mean-field approximations such as LSDA and GGA are enough to describe the ground-state 
properties with high accuracy; thus, we propose the LR-based DFT+U method, where the determined 
$U_\mathrm{eff}^\mathrm{LR}$ parameters are incorporated into only strongly-correlated $Y$ site, as a suitable 
methodology on a practical level for $L2_1$ Co-based full Heusler alloys.
For Co$_2$MnSi, our results are consistent with the experimental observations and superior to the standard GGA
calculation, particularly in terms of electronic and magnetic properties.
It is also indicated that Co$_2$MnSi is not HM but a highly spin-polarized ferromagnet.
Further investigations were carried out for the other ternary and quaternary
Co$_2$(Y,Mn)Si to explore the potential for HM ferromagnets.
The results showed that the Co$_2$(Ti,Mn)Si, Co$_2$(V,Mn)Si, and Co$_2$(Fe,Mn)Si compounds are expected 
to be HM materials when the composition of $Y$ element is appropriately selected.
The Co$_2$(Cr,Mn)Si does not show HM property at every composition, but a notable tendency is that the
high spin polarization is independent of the composition.
However, for using spintronics applications, 
Co$_2$(Fe$_{0.25}$,Mn$_{0.75}$)Si, in which the HM nature is consistent with the experimental
AMR study, is one of the most promising candidates because of the sizable HM gap in the minority state 
and the Fermi energy position being at almost the center of the gap.

\acknowledgments

The authors are grateful to K. Masuda, H. Sukegawa, and S. Mitani for critical comments and suggestions.
K. N. also thanks K. Nakamura, T. Oguchi, and M. Weinert for fruitful discussions.
This work was supported in part by Grants-in-Aid for Scientific Research (S) 
(Grant Numbers~JP16H06332 and JP17H06152) from the Japan Society for the Promotion of Science,
the ImPACT Program of Council for Science, Technology and Innovation, 
'Materials research by Information Integration Initiative ($Mi^2i$)' project of the
Support Program for Starting Up Innovation Hub from Japan Science and Technology Agency (JST),
and Center for Spintronics Research Network (CSRN), Osaka University.
The computations in this study were performed on a Numerical Materials Simulator at NIMS.

%... references


\begin{thebibliography}{99}
\bibitem{note_ulrt_alpha-size}
    In order to consider an effect of size of applied potential shift, 
    various values of $\mu_\mathrm{Co}$ were tested to compute the $U_\mathrm{eff}^\mathrm{LR(Co)}$,
    for example, $\mu_\mathrm{Co}=\pm0.06, \pm0.1, \pm0.5, \pm1.0$, and $\pm2.0$~eV.
    The obtained parameters are almost same with the difference less than 0.05~eV with an exception of
    $\mu_\mathrm{Co}=\pm2.0$~eV case, where the stationary self-consistent solution was not obtained
    because of the too large potential shift.
\bibitem{note_ulrt_Co2YSi}
    The LR calculations were also done for the $U_\mathrm{eff}^\mathrm{LR(Co)}$ in Co$_2Y$Si.
    Similar to the Co$_2$MnSi, large values of 6.536, 6.563, 6.674, and 5.782~eV are obtained for Co of Co$_2Y$Si, 
    where $Y$ is Ti, V, Cr, and Fe, respectively.
    Based on the discussion in Sec.~\ref{sec:CMS}, the $U_\mathrm{eff}^\mathrm{LR}$ values for Co is not used
    for the calculations in present work.
\end{thebibliography}
\end{document}